\documentclass[sigconf]{acmart}
\usepackage{amsmath,amsfonts}
\usepackage{algorithmic}
\usepackage{graphicx}
\usepackage{textcomp}
\usepackage{xcolor}
\usepackage{soul}

\usepackage[ruled, lined]{algorithm2e}
\SetKwRepeat{Do}{do}{while}
\makeatletter
\let\oldnl\nl% Store \nl in \oldnl
\newcommand{\nonl}{\renewcommand{\nl}{\let\nl\oldnl}}% Remove line number for one line
\makeatother
\newlength\mylen
\newcommand\myinput[1]{%
  \settowidth\mylen{\KwIn{}}%
  \setlength\hangindent{\mylen}%
  \hspace*{\mylen}#1\\}

\usepackage{array}

\newcommand{\code}[1]{\texttt{#1}}
    
\usepackage[caption=false]{subfig}
\newcommand{\rulesep}{\unskip\ \vrule\ }

\AtBeginDocument{%
  \providecommand\BibTeX{{%
    \normalfont B\kern-0.5em{\scshape i\kern-0.25em b}\kern-0.8em\TeX}}}

\setcopyright{acmcopyright}
\copyrightyear{2023}
\acmYear{2023}
\acmDOI{XXXXXXX.XXXXXXX}

\acmConference[Arxiv'23]{Make sure to enter the correct
  conference title from your rights confirmation emai}{Apr 16,
  2023}{USA}
\acmPrice{15.00}
\acmISBN{978-1-4503-XXXX-X/18/06}

\begin{document}

% TODO: REMOVE
\pagenumbering{arabic}

\title{Reclaimer: A Reinforcement Learning Approach to Dynamic Resource Allocation for Cloud Microservices\\ Paper Type: Regular}


\author{Quintin Fettes}
\email{qf731413@ohio.edu}
\affiliation{%
  \institution{Ohio University}
  \country{USA}
}
\author{Avinash Karanth}
\email{karanth@ohio.edu}
\affiliation{%
  \institution{Ohio University}
  \country{USA}
}
\author{Razvan Bunescu}
\email{rbunescu@uncc.edu}
\affiliation{%
  \institution{University of North Carolina at Charlotte}
  \country{USA}
}
\author{Brandon Beckwith}
\email{bbeckwi2@uncc.edu}
\affiliation{%
  \institution{University of North Carolina at Charlotte}
  \country{USA}
}
\author{Sreenivas Subramoney}
\email{sreenivas.subramoney@intel.com}
\affiliation{%
  \institution{Intel}
  \country{India}
}

\begin{abstract}
Many cloud applications are migrated from the monolithic model to a microservices framework in which hundreds of loosely-coupled microservices run concurrently, with significant benefits in terms of scalability, rapid development, modularity, and isolation. However, dependencies among microservices with uneven execution time may result in longer queues, idle resources, or Quality-of-Service (QoS) violations.

In this paper we introduce Reclaimer, a deep reinforcement learning model that adapts to runtime changes in the number and behavior of microservices in order to minimize CPU core allocation while meeting QoS requirements. When evaluated with two benchmark microservice-based applications, Reclaimer reduces the mean CPU core allocation by 38.4\% to 74.4\% relative to the industry-standard scaling solution, and by 27.5\% to 58.1\% relative to a current state-of-the art method.
\end{abstract}

\begin{CCSXML}
<ccs2012>
   <concept>
       <concept_id>10010520.10010521.10010537.10003100</concept_id>
       <concept_desc>Computer systems organization~Cloud computing</concept_desc>
       <concept_significance>500</concept_significance>
       </concept>
 </ccs2012>
\end{CCSXML}

\ccsdesc[500]{Computer systems organization~Cloud computing}

\keywords{microservices, reinforcement learning, resource allocation}



\maketitle

\section{Introduction}
% Describe the environment that has led to the problem
As cloud services grow in complexity, providers such as Uber \cite{gluck-introducing-2020}, Netflix \cite{mauro-microservices-2015}, and Google \cite{rzadca-autopilot-2020} are adopting a microservices model, wherein services are comprised of dozens or hundreds of independent, loosely-coupled microservices. The microservices model offers benefits in terms of scalability, modularity, isolation and maintainability \cite{gan-open-source-2019, jamshidi-microservices-2018}. However, cloud services have a strict Quality-of-Service (QoS) requirement -- the maximum allowable $99^{th}$ percentile end-to-end latency \cite{mauro-microservices-2015, satnic-amazon-nodate, esposito-challenges-2016}. 

There are several reasons why meeting the stringent QoS constraints in a microservices model becomes challenging: (i) microservices have constantly fluctuating workloads that change with the number and types of requests, and spiking workloads increase latency as queues are filled, (ii) requests must interact with several microservices (sometimes sequentially) in order to be satisfied, leading to increased delays, and (iii) individual microservices can be updated, added, and removed throughout execution, adding uncertainty to their resource requirements. For example, a social media service might update its content recommendation algorithm, or add a new microservice which automatically tags users when photos are uploaded. 
% These challenges can be addressed by allocating sufficient hardware resources such as CPU cores, caches, network bandwidth, memory bandwidth, and I/O priority. 
When a microservice is allocated insufficient resources for the workload at hand, it cannot satisfy all user requests in a timely fashion, which in turn, causes the delay to propagate to downstream microservices, ultimately resulting in QoS violations \cite{gan-sage-2021, gan-seer-2019, zhang-sinan-2021}. QoS violations define a bad user-experience, and as a result, have a negative monetary impact to the service providers. For example, Amazon found that a $100$ ms increase in latency led to a $1\%$ reduction in sales \cite{linden-geeking-2006}. Thus, it is crucial to develop resource management policies which do not over-allocate and waste hardware resources, but still avoid costly QoS violations.

Microservices are often over-provisioned with computational resources such that a rapid increase in the number of service requests does not result in QoS violations. As a result, underutilized servers waste idle hardware resources \cite{zhang-sinan-2021, lo-heracles-2015, qiu-firm-2020}. 
Performance debugging methods \cite{gan-sage-2021, gan-seer-2019} identify the root cause of QoS violations. When the culprit microservice and resource are identified, the resource allocation is increased to prevent further QoS violations. Autoscaling approaches utilize machine learning (ML) techniques \cite{gupta-deeprecsys-2020, mao-neural-2017, park-graf-2021} or manually engineered policies \cite{lo-heracles-2015, patel-clite-2020} to mitigate QoS violations by scaling the allocation of resources such as CPUs, memory, caches, I/O channels, and network links.

% Brief description of the key points in your method, how your method fills the gap and brief preview of results
Prior approaches to  allocate resources for microservices under QoS constraints suffer from two limitations. First, reactive approaches excel at identifying the cause of violations and adjusting resource allocations, but are limited by their inability to predict and prevent an upcoming violation. This is significant because poor resource allocations can be difficult to recover from, and have long-term effects on the tail latency. Second, most existing algorithms cannot match the dynamic nature of the microservices themselves. Manually engineered algorithms may require new, hand-tuned heuristics when microservices receive significant updates, and parameterized ML approaches often have a built-in assumption of stationarity and thus cannot adapt to a variable number of microservices without retraining all parameters from scratch. Other approaches are also being constrained to specific applications, or relying on special instrumentation of the applications themselves to provide tracing data.

To address these shortcomings, we propose \textsc{Reclaimer}, an approach based on Deep Reinforcement Learning (DRL) that avoids QoS violations altogether by proactively adjusting CPU core allocations to individual microservices in an elegant, end-to-end fashion without the use of multiple, highly engineered models. To make core allocation decisions, Reclaimer uses local feature information collected from the Docker interface to the Linux cgroups API. Critically, Reclaimer re-uses parameters to process the input information for each microservice. This allows Reclaimer to be dynamic to a variable number of microservices with a uniform policy architecture, and learn a policy which can rapidly adapt to new microservices and updates to existing microservices.
% Finally, Reclaimer utilizes the off-policy algorithm Soft Actor-Critic (SAC) to learn an effective policy for core allocation.

% contributions
To the best of our knowledge, Reclaimer is the first proactive, data-driven core allocation algorithm that automatically adjusts to common workload fluctuations, such as the number of microservices being co-located, updates to individual microservices, updates to the dependencies among microservices and the number of available CPU cores in the system. Because Reclaimer collects input information without special instrumentation and avoids per-application tuning, it can be used on existing microservice architectures with marginal design overhead. 
The major contributions of this work are as follows:
\begin{itemize}
     \item \textbf{Proactive CPU allocation:} Reclaimer collects runtime information every one second, and recommends a new core allocation using the most up-to-date information. Indicators of potential future QoS violations are effectively captured by the proposed features, and Reclaimer avoids the QoS violations.
     \item \textbf{Online Training:} Reclaimer collects, stores, and updates information in a fully online fashion. Online learning causes little interference with microservice processing. Because the underlying RL algorithm is SAC \cite{haarnoja-soft-2018}, Reclaimer can make efficient use of previously collected experience to keep performing updates and adjust to workload fluctuations which occur in real microservice deployments.
     \item \textbf{Adaptability:} By re-using parameters for all microservices, Reclaimer is able to effectively learn a policy for a variable number of microservices, which may change or be replaced entirely. Reclaimer does not require new, randomly initialized parameters to learn a policy for new microservices, and can rapidly adapt to significant changes in the service.
 \end{itemize}

% brief summary of results \cite{noauthor-aws-nodate}
Reclaimer outperforms the industry-standard AutoScale  and Sinan \cite{zhang-sinan-2021}, a state-of-the-art resource allocation method. On the two applications from the DeathStarBench microservice benchmark suite on which Sinan was tested, Reclaimer met the QoS requirement $100\%$ of the time, while reducing mean CPU core allocation by  $38.4\% - 74.4\%$ relative to Autoscale, and by $27.5\% - 58.1\%$ relative to Sinan. Additionally, we utilize transfer learning to demonstrate that the policy learned by Reclaimer generalizes to extreme changes in the microservice dependency graph by showing that, for Social Media, a policy pretrained on Hotel Reservation is able to learn over $2\times$ as quickly as a randomly initialized policy. 
% Finally, we utilize Shapely Additive Explanations (SHAP) \cite{lundberg-unified-2017} to interpret the policy learned by Reclaimer. Specifically, we examine the relative impact of the input features on the resource allocation decisions made for two distinct microservices, and explain why the learned policy makes specific scaling decisions. We also demonstrate that the learned policy is able to learn unique characteristics of individual microservices, despite sharing parameters for each microservice.

\section{Problem Overview and RL}
\label{sec:overview}
The objective of this resource allocation problem is to meet the end-to-end latency requirement of microservice-based cloud services while minimizing CPU core allocation. Given the set of microservices $M$, the maximum allowable $99^{th}$ percentile tail latency $QoS$, the observed tail latency $L_t$ at time $t$, and the core allocation $c_t(m)$ for a microservice $m$, the objective is to minimize the total core allocation over time $\sum\limits_{t=0}^{\infty} \sum\limits_{m \in M} c_t(m)$ subject to $L_t < QoS$, $\forall t$. Over-allocating CPU cores results in idle, wasted cores. On the other hand, allocating an insufficient number of cores to a microservice can lead to rapidly filling its queue with pending requests, which will increase latency. This can have a cascading effect on the latency of downstream microservices, ultimately leading to an increase in the tail latency for the service as a whole. 
%To offset the increase in tail latency, higher core allocations to the microservice, while beneficial, will take longer to empty the queues and return the tail-latency to an acceptable level. 

\begin{figure}[!tb]
\centering
\includegraphics[width=0.9\columnwidth]{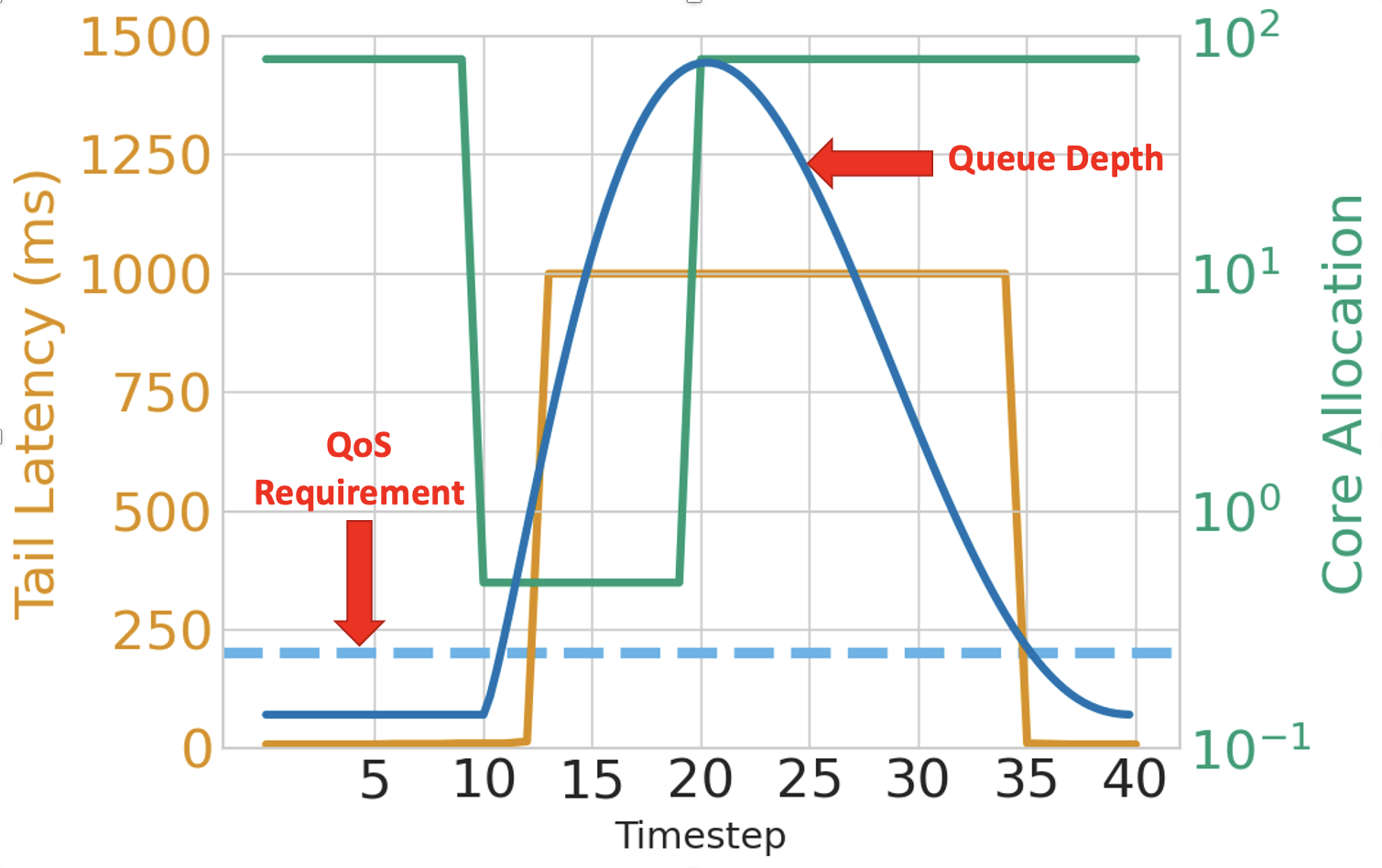}
\caption{This example demonstrates the long-term effects of sub-optimal resource allocation decisions. The workload is set to its maximum value of $4000$ users. The green line and axis correspond to the core allocation for each microservice, the yellow line and axis correspond to the tail latency, and the blue line is a representation of the queue depth for each microservice. The QoS requirement is represented by the horizontal blue line. Initially, all microservices are given the maximum possible core allocation. At timestep $10$, each microservice is given an insufficient core allocation, $0.5$ cores. At timestep $20$, the maximum core allocation is restored to each microservice.}
\label{fig:queuingEffect}
\end{figure}

This work considers two microservices benchmark applications from the DeathStarBench \cite{gan-open-source-2019} benchmark suite that were used by prior work, Sinan \cite{zhang-sinan-2021}. Social Media is implemented using 28 microservices, whereas
%Users can post messages with text, media, links, and references to other users, utilize image and text filter services, and read posts from other users. 
Hotel Reservation is implemented using 15 microservices.
%, and enables users to search for hotels, place reservations, and receive recommendations. 
To simulate real workloads, we utilize the Locust \cite{noauthor-locust-nodate} workload generation tool. Figure \ref{fig:queuingEffect} shows the detrimental effects of under-allocation for Hotel Reservation. At timestep $10$, all microservices are allocated $0.5$ CPU cores each. After three timesteps (timestep $13$ onwards), the queues fill and the service begins consistently violating its QoS requirement. After $10$ seconds of insufficient allocations, at timestep $20$ the microservices are all given their maximum possible core allocation. In this example, it takes 15 seconds (timesteps $20$ to $35$) for queues to empty, allowing the service to recover from this extended period of insufficient allocations. This examples demonstrates the need to proactively adapt to a rapidly changing microservice environment, for which an approach based on Deep Reinforcement Learning (DRL) is a natural choice.

% Figure \ref{fig:queuingEffect} demonstrates that a successful algorithm must determine the impact of core allocation decisions which have delayed and long-term consequences for the service latency. Moreover, the algorithm must be able to co-locate an arbitrary number of microservices on an arbitrary number of CPU cores, and adapt to a rapidly changing microservices. To meet these requirements, an approach based on Deep Reinforcement Learning (DRL) is a natural choice. DRL is formulated to find policies which optimize for long-term objectives, where actions may have delayed and long-term effects \cite{sutton-reinforcement-2018}. Additionally, DRL does not assume the underlying problem is stationary \cite{sutton-reinforcement-2018}, which allows the service, individual microservices, and available resources to change over time. Finally, the problem can be formulated such that allocation decisions are proactive, and information gained from special instrumentation of the application is unnecessary. Altogether, DRL provides a framework to learn a core allocation policy which matches the complex and dynamic nature of the microservice model itself.

\subsection{Reinforcement Learning (RL)}

RL is a type of machine learning in which the agent aims to find an optimal policy for selecting actions. The selected actions influence the agent's environment, and after each action the environment provides a scalar reward describing progress toward a goal. The prevailing problem framework for RL is the Markov Decision Process (MDP), represented as a tuple $(S, A, R, P, \gamma)$ where $S$ is the set of all states $s$, $A$ is the set of all possible actions $a$ the agent can take, $R$ is the reward function, $P(s',r |s, a)$ is the dynamics model of the environment describing the probability of transitioning into a new state $s'$ and observing a reward $r$ when action $a$ is taken from state $s$, and $\gamma$ is the discount factor. The agent aims to learn an optimal policy $\pi^* : S \rightarrow A$ that maps states to actions such that the long term expected reward is maximized. 

% The problem of finding an optimal policy $\pi^*$ can be reformulated as learning an action-value function $Q^*(s,a)$ that maximizes the expected cumulative reward:
% \begin{equation}
%   Q^*(s_t, a_t) = \max\limits_{\pi} \mathbb{E}_P\big[r_{t+1} + \gamma r_{t+2} + \gamma^2 r_{t+3} + ... | s_t, a_t \sim \pi \big]
% \end{equation}
% which is the expected sum of rewards $r_i$, discounted at each timestep $i$ by a positive discounting factor $\gamma \leq 1$, when actions are taken according to policy $\pi$, maximized over all possible policies \cite{watkins_q-learning_1992}. It can be shown that the optimal policy is $\pi^* = \argmax\limits_{a}Q^*(s,a)$ \cite{sutton_reinforcement_2018}.

% An underlying assumption of the MDP is that the input states $s$ contain all relevant information for selecting the action which maximizes the expected long-term reward. However, in some cases, the input given to the agent does not satisfy this constraint, which corresponds to the Partially Observable Markov Decision Process (POMDP) problem framework. In this framework, the agent receives observations $o \in O$, where one observation can correspond to one or more states $s \in S$ in the underlying MDP. 
%One special case of the POMDP occurs when an observation lacks certain temporal information which, if it were included, would make the POMDP a fully-observable MDP. 
%Recalimer uses $k=5$ stacked observations to ensure the environment is fully-observable by including information on the change in workload over time.

In cases where the state space, the action space, or both are too large to explicitly represent a state-action value table, deep reinforcement learning algorithms such as Deep Q-Networks \cite{mnih-human-level-2015}, Asynchronous Advantage Actor Critic \cite{mnih-asynchronous-2016}, Soft Actor-Critic (SAC) \cite{haarnoja-soft-2018} can be used to learn a representation of the action-value function, the policy function, or both via a function approximation model, e.g. a neural network.

% If the actions are chosen such that, in the limit, all actions are taken in all states a sufficient number of times, many RL algorithms are guaranteed to converge to the optimal action-value function. This poses a central conflict between exploration and exploitation. Exploitation will allow the agent to maximize its return in the short term. However, more exploration may allow the agent to better estimate the true value function and lead to a greater long-term return \cite{sutton-reinforcement-2018}. One modern approach to improve exploration strategies is achieved by SAC through the use of the maximum entropy RL framework \cite{haarnoja-soft-2018}.

\section{Reclaimer}
\label{sec:system_overview}
\begin{figure}[!tb]
\centering
\includegraphics[width=\columnwidth]{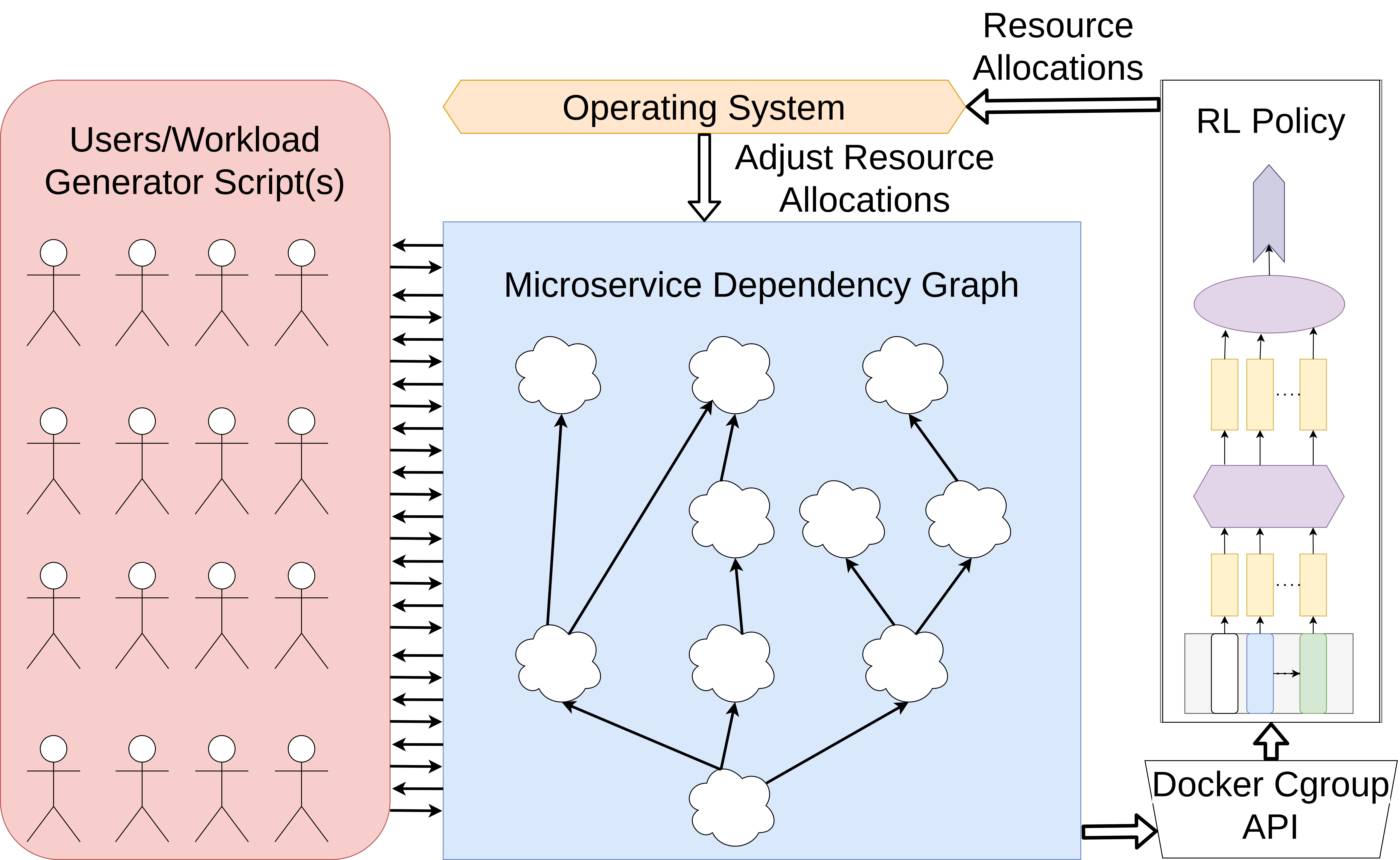}
\caption{The overall design of Reclaimer. Users (or a workload generation script) interact with a microservice-based service. Information about the performance and workload of individual microservices is collected via the Linux cgroups API. The Reinforcement Learning policy uses this information to determine CPU core allocations for each microservice. Core allocations are passed to the operating system, which then makes the adjustments.}
\label{fig:systemDesign}
\end{figure}

The overall design of Reclaimer is shown in Figure \ref{fig:systemDesign}. Users or workload simulation scripts interact directly with the service, a DRL policy collects information from the Docker interface to the Linux cgroups API, gives core allocations to the operating system, and the operating system enforces CPU utilization limits on the individual microservices. Our experiments are performed on a server with dual socket, Intel Xeon Gold 6230N 20-core, 40-thread CPUs, and two Tesla V100 GPUs. The same server is utilized to run the microservices and execute the DRL workload simultaneously. While there is some interference between the workloads on the CPU, much of the DRL training is performed on the GPU, which is not utilized by the microservices. Overhead can be quantified by the inference time for the policy network: for Social Media inference takes $2.7$ms, and for Hotel Reservation inference takes $2.3$ms. 
%At production scale, the DRL workload could be completely offloaded to a low-load server. Furthermore, the off-policy nature of the algorithm allows the training data to be used for an arbitrary number of updates, regardless of the current policy. Thus, updates could be intelligently scheduled to occur only off-peak hours at virtually no performance penalty.
To achieve the best performance, Reclaimer should observe and allocate resources for all microservices running on a CPU. The MDP and the model architecture do not assume that all microservices are part of the same application. Microservices that are not managed by Reclaimer, such as the Jaeger microservice in our experiments, are considered as part of the RL environment.

The features shown in Table \ref{tab:MSFeatures} are collected for each microservice and capture runtime statistics relating to resource utilization and performance: communication features describe communication among microservices, memory features capture information about on and off-chip memory utilization, CPU features describe CPU utilization and allocation, IO features describe accesses to IO devices, latency features capture the current latency distribution, and request service features describe the current workload, in terms of both requests and failures per second. Most feature categories are collected by utilizing Docker's interface to the Linux cgroups API. The remaining feature categories - latency and request service - are collected from the Locust \cite{noauthor-locust-nodate} workload generation tool. In a production setting, the features collected via Locust would be measured by a lightweight process which forwards requests to the downstream service. Timestamp information would be sent by the client to compute network latency, latency within the application would be measured by examining the request and response timestamps in the new process, and request counts would be tracked with counters.
%Locust is used to simulate real users \cite{noauthor-locust-nodate} interacting with the cloud service benchmarks. 
In all cases, feature collection is accomplished by reading from files stored on disk. By utilizing the cgroups API and Locust, Reclaimer does not require special implementation via a tracing service to collect information like queue depths. 
% This allows Reclaimer to be more general, and work with existing microservice-based services without requiring extensive development overhead. %However, due to limitations within Docker, feature information can only be collected every 1s. Thus, Reclaimer is limited in making core allocation decisions every 1s. 

During training and testing, the microservice core allocations are manipulated via calls to \code{docker update --cpus=core\_allocation}. Next, Docker restricts the CPU utilization of all processes in a Docker container by calculating appropriate values for the \textit{CPU quota} and \textit{CPU period} parameters of each process group \cite{noauthor-runtime-2021} for the Linux completely fair scheduler \cite{noauthor-cfs-nodate}. This limits the core utilization of each microservice, while still allowing them to be scheduled to any CPU core in the system. Feature information is collected once every second due to limitations of Docker, so collecting data is relatively time-consuming when compared to common RL benchmarks \cite{bellemare-arcade-2013, coumans-pybullet-2016}. 
%The feature, action selection, and latency data used to compute the reward is stored in a database. 
This is addressed by the off-policy training procedure used by SAC, which allows the agent to efficiently re-use collected data to perform many updates. 
%The entire service and the mechanisms to collect features, extract rewards, and execute actions are wrapped as an OpenAI gym \cite{brockman-openai-2016} environment for easy use with existing DRL implementations.

\begin{table}[!tb]
\centering
\caption{Input features for each microservice, measured over a 1-second feature collection interval.}
 \begin{tabular}{|| c | p{15em} ||} 
 \hline
 Feature Category & Features \\ [0.5ex] 
 \hline\hline
 Communication & Packets Received, Packets Sent, Bytes Received, Bytes Sent\\ 
 \hline
 Memory & Resident Set Size (RSS), Cache Memory, Page Faults\\
 \hline
 CPU & CPU Time, Maximum Core allocation \\
 \hline
 IO & IO Bytes, IO Services\\
 \hline
 Latency & $50^{th}$, $66^{th}$, $75^{th}$, $80^{th}$, $90^{th}$, $95^{th}$, $98^{th}$, $99^{th}$, $99.9^{th}$, $99.99^{th}$, $99.999^{th}$, $100^{th}$ Percentile End-to-End Latency, QoS Requirement\\
 \hline
 Request Service & Requests Per Second, Failures Per Second\\
 \hline
 Other & Unique Identifier, Previously-selected Action\\
  \hline 
 \end{tabular}
\label{tab:MSFeatures}
\end{table}

\subsection{Markov Decision Process Formulation}

\subsubsection{Observations and States} 

\begin{figure}[!t]
\centering
\includegraphics[width=0.9\columnwidth]{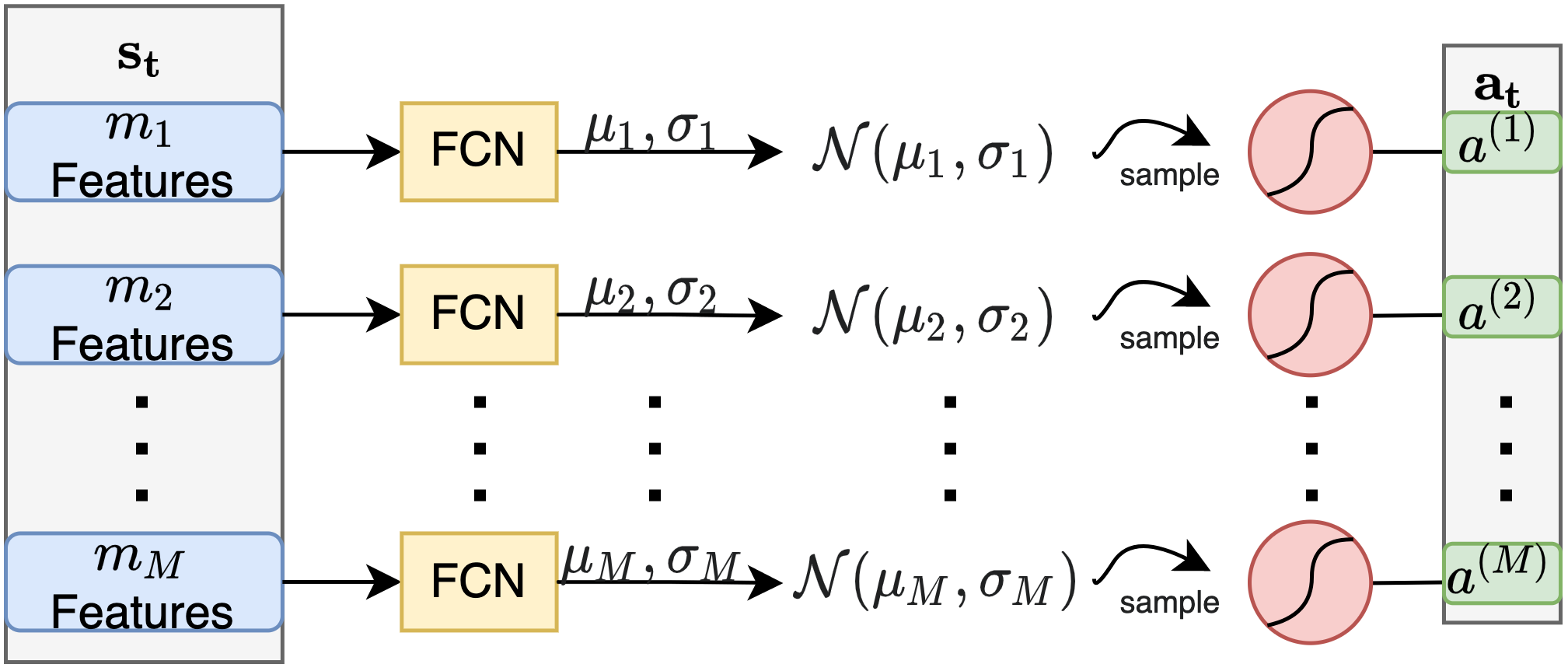}
\caption{The policy function neural network architecture.}
\label{fig:MSPolicyNetwork}
\end{figure}

\begin{figure*}[!t]
\centering
\subfloat[]{\includegraphics[width=0.49\textwidth]{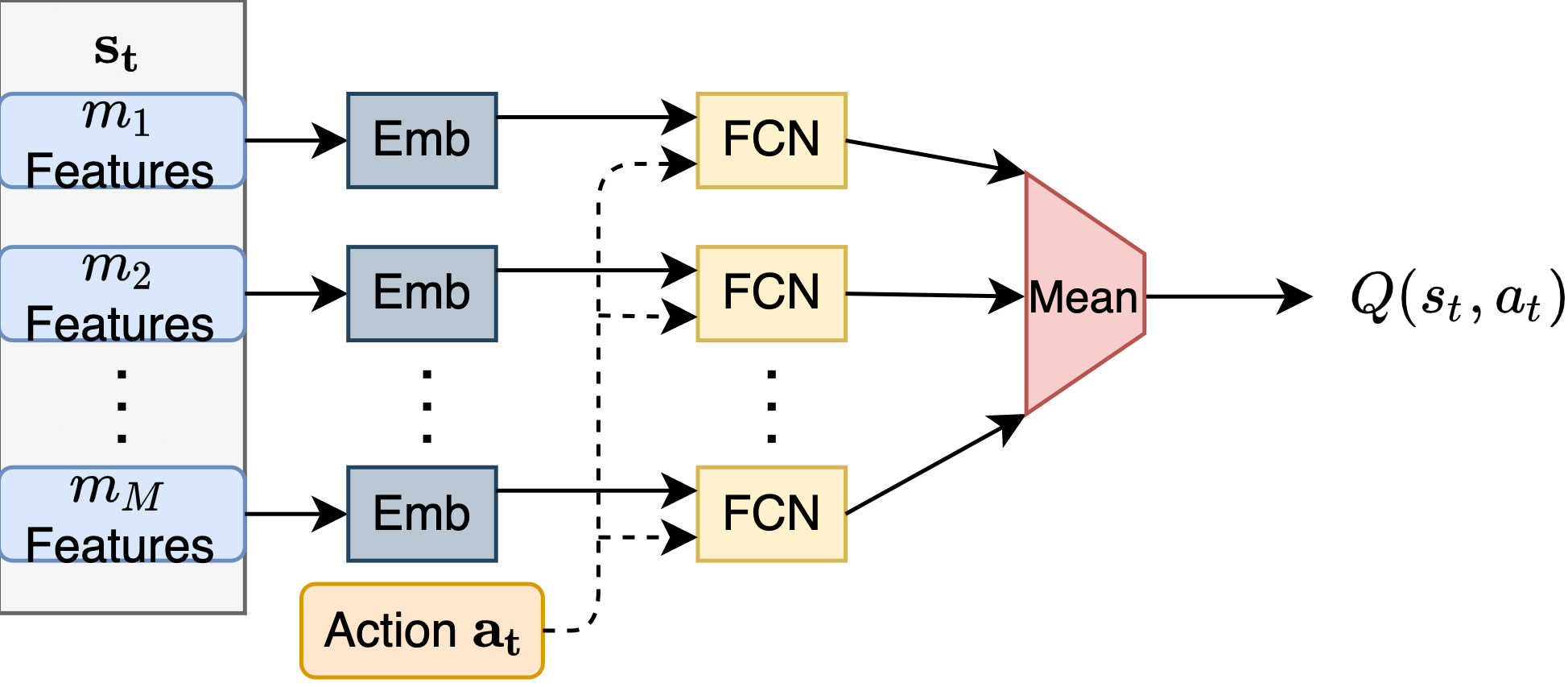}%
\label{fig:ValueNetwork}}
\hfil
\hfil
\rulesep
\subfloat[]{\includegraphics[width=0.46\textwidth]{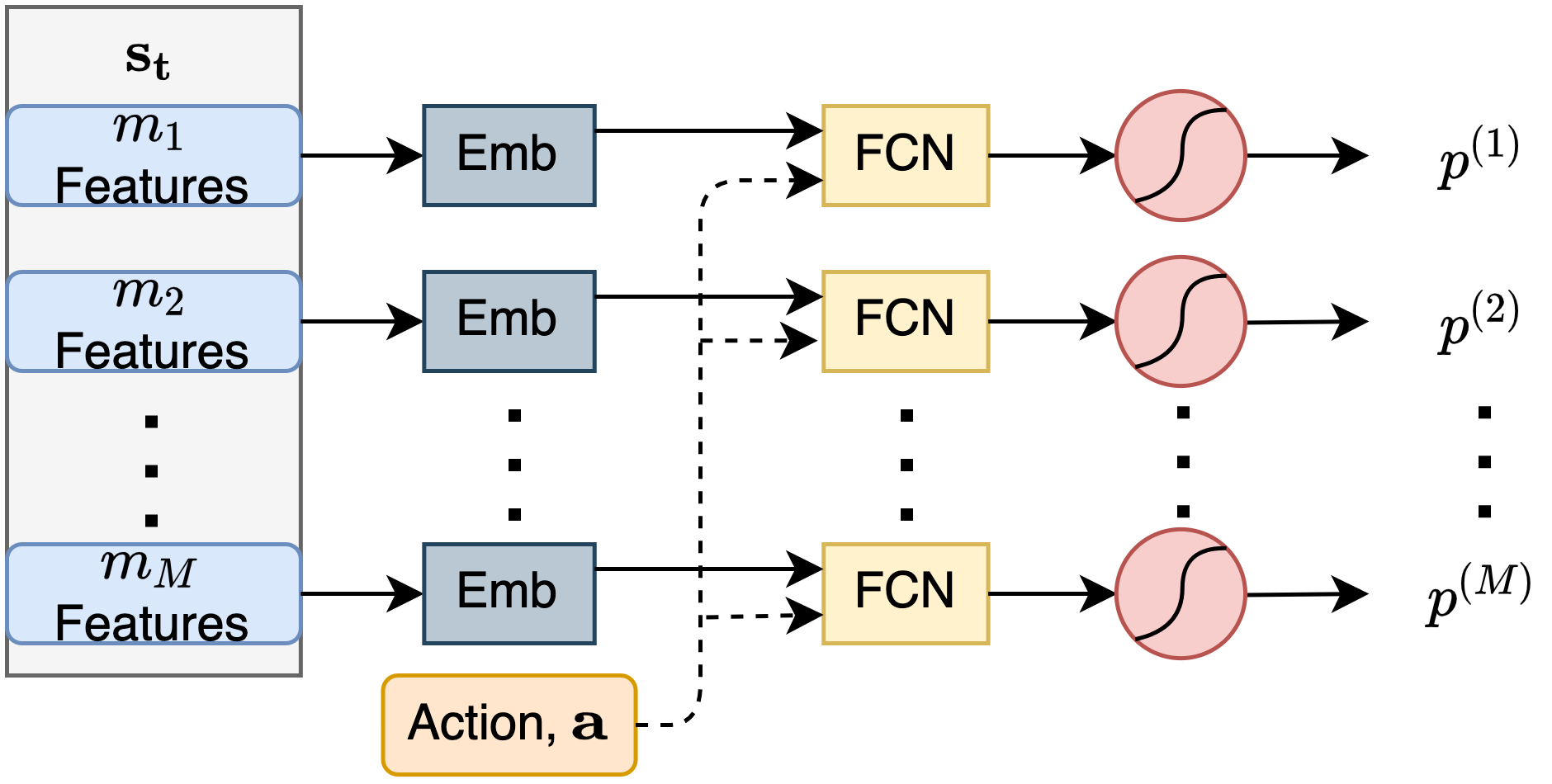}%
\label{fig:InfoGain}}
\hfil
\caption{\textbf{(a)} shows the neural network architecture to compute the state-action value function given a state $s_t$ and the selected action $a_t$. \textbf{(b)} shows a neural network architecture to predict the probability of meeting the QoS requirement given a candidate action $a$. Probabilities are computed independently for each microservice.}
\label{fig:MSAdditionalNetworks}
\end{figure*}

For the set $M$ of all microservices to be managed by Reclaimer, the observation space is composed of one vector for each microservice $m \in M$. For notational simplicity, we will also use $M$ to denote the size of the set $M$. Thus, if there are $n$ elements in each feature vector, the observation matrix is defined such that $o \in \mathbb{R}^{M\times n}$, $\forall o \in O$.  Note that latency features and request service features shown in  Table \ref{tab:MSFeatures} measure service-level statistics, and are identical for all microservices at each timestep. The unique identifier is a one-hot vector for each microservice.

To compute the state matrix, Reclaimer concatenates the $k$ most recent observations along the second dimension to form the 2-dimensional state matrix $s \in \mathbb{R}^{M \times (kn)}$. By concatenating consecutive observations, the model is given temporal information describing the change in the workload over the most recent $k$ timesteps. 
%Altogether, the state matrix requires low-overhead to collect and does not require special instrumentation of the application. Each input feature vector is standardized before being used as input to the DRL algorithm. 
One global mean and one global variance vector are computed and used to standardize the input for each microservice. %(i.e. the same mean vector and variance vector are used to standardize the unique microservices x and y).

\subsubsection{Actions}
The action vector defines a percentage of the total cores in the system allocated to each microservice. In this formulation, sub-core allocations are possible and microservices can be co-located on the same core. Thus, the action vector at time $t$ is defined as $\mathbf{a}_t \in \mathbb{R}^{M}$, where $a_t^{(i)} \in [0, 1]$ $\forall$ $i \in [0, M-1]$. The true core allocation for microservice $m_i$ is defined as $a_t^{(i)} u^{(i)}$ where $u^{(i)}$ is a user-defined upper bound on the core allocation for the microservice $m_i$. Individually defining maximum core allocations for microservices is common, and the same approach is taken by Sinan \cite{zhang-sinan-2021}. For Reclaimer, manually defining maximum core allocation allows the model to more easily learn core allocations for microservices which require a significantly different number of cores on average, but they are not necessary for convergence. In the absence of prior knowledge concerning upper bounds on core utilization, the maximum available cores could be used for all values of $u_i$ at the cost of increased training time.

\subsubsection{Rewards} 
The reward function is shown below:

\[ r_t(a_{t-1}) =
\begin{cases} 
  -1 & L_t > QoS \\
  \alpha(1 - \frac{1}{Z} \mathbf{u}^T \mathbf{a}_{t-1})  & L_t \leq QoS
\end{cases}
\label{eq:reward}
\]

\noindent where $r_t$ is the reward at timestep $t$, $L_t$ is the $99^{th}$ percentile latency over the 1-second timestep, $QoS$ is the end-to-end latency requirement for the service, $\alpha$ is a tunable hyperparameter, $\mathbf{a_{t-1}}$ is the CPU allocation vector at timestep $t-1$, $\mathbf{u}$ is the vector of per-microservice core caps, and $Z = \sum_{i=1}^{M} u^{(i)}$ is a normalization factor. First, $\alpha \geq 0$ is a tuneable hyperparameter to allow the designer to control the emphasis the agent places on meeting the QoS requirement. Because $\alpha$ is a positive weight on the positive reward provided by saving CPU cores, larger values of $\alpha$ result in an agent which opts to save more resources, and is more tolerant of occasional QoS violations. Similarly, smaller values of $\alpha$ make saving CPU cores less rewarding, and result in an agent which is more risk aversive, and will opt to output higher core allocations to avoid QoS violations. $\alpha$ was selected by doing a hyperparameter search over the values [0.1, 1.0, 10.0]. $\alpha=1.0$ yielded models which avoided QoS violations while still achieving good core allocation results on both benchmark applications. The reward at timestep $t$ is explicitly a function of the action at timestep $t-1$. When the $99^{th}$ percentile latency exceeds the QoS requirement, the reward is $-1$, the worst possible reward, which incentivizes the agent to meet the QoS requirement. 
%When the agent meets the QoS requirement but utilizes all cores, the reward will be 0.
While meeting the QoS requirement, the reward approaches $\alpha$ as the agent allocates fewer cores. Thus, the agent is incentivized to meet the QoS requirement, but also to allocate as few cores as possible. 

%\subsection{Soft Actor Critic (SAC)}

% Reclaimer utilizes SAC \cite{haarnoja-soft-2018} to learn the core allocation policy. SAC updates the parameters of a policy neural network in the direction that maximizes the current estimate of the soft action-value function. %This is desirable because SAC is compatible with continuous actions spaces (e.g. The $[0, 1]$ interval Reclaimer uses for CPU allocations) and SAC utilizes the maximum entropy framework to achieve a more principled method of exploration than simpler approaches such as $\epsilon$-greedy policies. 
%SAC is compatible with continuous action spaces, explores efficiently by utilizing the maximum entropy framework, and the capability to reuse samples collected over long simulations is crucial because experience is expensive to collect for the core-allocation problem. Note that Reclaimer uses a variant of SAC which automatically learns the value of the entropy coefficient \cite{haarnoja-soft-2019}. For a full description of the algorithm, the reader is referred to \cite{haarnoja-soft-2018}.

\subsection{Reinforcement Learning for Core Allocation}
Reclaimer utilizes Soft Actor Critic (SAC) \cite{haarnoja-soft-2018} to learn the core allocation policy. SAC is compatible with continuous action spaces, off-policy learning, and offline learning. Offline learning enables SAC to reuse samples collected over long simulations, which is very important for the core-allocation problem because experience is expensive to collect. The SAC algorithm for learning a core allocation policy via offline data is shown in Algorithm \ref{alg:SoftActorCritic}.

For each microservice $m \in M$, an action scalar $a^{(i)}_t$ is produced at each timestep, which represents the percentage of the maximum possible cores to allocate to a microservice. The action vector $a \in \mathbb{R}^M$ is computed via the policy neural network shown in Figure \ref{fig:MSPolicyNetwork}. The first component of the policy network is a sequence of $7$ fully-connected layers, each with $256$ hidden units. Figure \ref{fig:MSPolicyNetwork} represents the policy network as a neural network which processes a $M\times n$ input matrix and produces an action vector of length $M$ by sharing parameters across fully-connected layers. An alternative representation of the policy network is a fully connected network which process $M$ input vectors of length $n$ at each timestep to produce a scalar action for each input vector, then forms the action vector by concatenating each scalar into a vector of length $M$. This alternative representation makes it clear how the network would adapt to a variable number of microservices: Reclaimer simply uses the fully connected network to process $M\pm k$ vectors of length $n$ when adding or removing $k$ microservices. However, the input layer requires 1 unique parameter per hidden unit to process the unique identifier feature for each microservice. With the exception of the parameters for the unique identifier feature, \textit{the same parameters are used to process the input for each microservice}. As a result, for any single microservice $\sim 99.95\%$ of parameters are re-used, which allows the network to perform well when processing a variable number of input microservices after only a relatively small amount of fine-tuning (more details in Section \ref{sec:transferLearning}).

For each microservice, one fully-connected layer produces a mean scalar $\mu_i$, and a second fully-connected layer produces a standard deviation scalar $\sigma_i$ that is clipped to the range $[-20, 2]$ and exponentiated to ensure it is positive. Each mean and variance pair is used to parameterize a Normal distribution $\mathcal{N}(\mu_i, \sigma_i^2)$, $\forall i \in M$. During training, the action scalar for each microservice is sampled from the normal distribution, whereas during evaluation the mean is used for the action. Finally, the action scalars are used as input to the logistic sigmoid function so that they can be interpreted as a percentage. Recall that for each microservice, core allocation is $a_t^{(i)} \times u^{(i)}$, where $u^{(i)}$ is the maximum possible core allocation for the microservice $m_i$. 

The structure of the value network is shown in Figure \ref{fig:ValueNetwork}. The action for each microservice is appended to the output of the second fully connected layer, and the resultant vector is used as input for a sequence of $5$ fully-connected layers with $256$ hidden units each. Additionally, the final feedforward network produces a single scalar for each microservice.
%instead of a mean and variance scalar. 
The mean of the scalars for all microservices is used as the action-value estimate $Q(s_t, a_t)$. By using the mean, the magnitude of the output is independent of the number of input microservices. Thus, similar to the policy network, this value network can handle a variable number of microservices.

Finally, a model which estimates the probability of meeting the QoS requirement at each of the next $5$ timesteps is shown in Figure \ref{fig:InfoGain}. Like the policy and value networks, this network uses the same set of parameters to process each microservice. One probability is produced for each microservice $m$, which represents the probability of meeting the QoS requirement at each of the next $5$ timesteps if $m$ is assigned action $a^{(m)}$. This network is used to encourage the agent to explore more efficiently in the early stages of learning (discussed below).
%This network is used to collect initial training data by querying a set of candidate actions (discussed below), and selecting the action which allocates the fewest cores and will not cause a QoS violation with at least $80\%$ certainty. This is similar to the data collection scheme used in \cite{zhang-sinan-2021}. However, their work modeled the probability of a QoS violation as a Bernoulli distribution using historical data, did not learn the probability using a machine learning method, and attempted to maximize information gain while collecting samples. 

\section{Training and Evaluation}
\label{sec:training}

\LinesNumbered
\begin{algorithm}[!ht]
    \KwData{A microservice-based cloud service application}
    \KwIn{Initial policy network parameters $\theta$}
    \nonl\myinput{Initial action-value network parameters $\phi_1$, $\phi_2$}
    \nonl\myinput{Initial entropy coefficient $\eta$}
    \nonl\myinput{Replay memory buffer $D$ with capacity $N$}
    \KwOut{Updated parameters $\phi_1$, $\phi_2$, and $\theta$}
    
    \vspace{1em}
    
    Set target parameters $\hat{\phi}_{1} \leftarrow \phi_1$, $\hat{\phi}_{2} \leftarrow \phi_2$\\
    
    % \Do{not converged}{
    %     Observe state $s$ and select action $a\sim\pi_{\theta}(\cdot | s)$\\
    %     Execute $a$ in the environment\\
    %     Observe next state $s'$, reward $r$, and done signal $d$\\
    %     Store ($s, a, r, s', d$) in replay buffer $D$\\
    %     if $s'$ is terminal, reset the environment state\\
        
        \For{j in range(number of updates)}{
            Randomly sample a batch of transitions $B = \{(s, a, r, s', d)\}$ from $D$\\
            
            Compute targets for the $Q$-functions:\\
            $H = -\log\pi_{\theta}(\Tilde{a}'|s')$\\
            \nonl $y(r, s', d) = r + \gamma (1-d)[\min\limits_{i=1,2}Q_{\hat{\phi}_{i}}(s', \Tilde{a}') + \eta H] $  \\ 
            \nonl where $\Tilde{a}' \sim \pi_{\theta}(\cdot| s')$\\
            
            Update $\phi_1, \phi_2$ by 1 step of gradient descent:\\
            \nonl $\nabla_{\phi_i} \frac{1}{|B|} \sum\limits_{(s, a, r, s', d) \in B} (Q_{\phi_i}(s, a) - y(r, s', d))^2$, for $i=1, 2$\\
            
            Update $\theta$ by 1 step of gradient ascent:\\
            \nonl $H = -\log \pi_{\theta}(\Tilde{a}_{\theta}(s) | s)$\\
            \nonl$\nabla_{\theta} \frac{1}{|B|} \sum\limits_{(s, a, r, s', d) \in B} [\min\limits_{i=1,2}Q_{\phi_{i}}(s, \Tilde{a}_{\theta}(s)) + \eta H]$,\\
            \nonl where $\Tilde{a}_{\theta}(s)$ is a sample from $\pi_{\theta}(\cdot | s)$ which is\\
            \nonl differentiable w.r.t. $\theta$ via the reparameterization trick\\
            
            Update $log\eta$ by 1 step of gradient ascent:\\
            \nonl $\nabla_{\eta} \frac{1}{|B|} \sum\limits_{(s, a, r, s', d) \in B} log\eta * (log\pi(a, s | \theta) + t_{H})$,\\
            
            Perform Polyak update of target networks:\\
            \nonl $\hat{\phi}_{i} \leftarrow \rho \hat{\phi}_{i} + (1-\rho)\phi_i$, for $i = 1, 2$\\
        }
    % }
\caption{Offline Soft Actor-Critic \cite{haarnoja-soft-2018, haarnoja-soft-2019, achiam-spinning-2018} }
\label{alg:SoftActorCritic}
\end{algorithm}

\begin{table}[!tb]
\centering
\caption{Table of hyperparameters for Reclaimer.}
 \begin{tabular}{|| c | c | c ||} 
 \hline
 \textbf{Hyperparameter} & \textbf{Symbol} & \textbf{Value} \\ [0.5ex] 
 \hline\hline
 QoS Target (Social) & $QoS$ & $500$\\
 \hline
 QoS Target (Hotel) & $QoS$ & $200$\\
 \hline
 QoS Modifier & $\alpha$ & $1.0$\\
 \hline 
 FC Layers & - & $7$\\
 \hline
 FF Hidden Units & - & $256$\\
 \hline
 Min/Max Users (Social) & $U_{min}/U_{max}$ & $[20, 200]$\\
 \hline
 Min/Max Users (Hotel) & $U_{min}/U_{max}$ & $[500, 3500]$\\
 \hline
 Individual Experiment Time & $e_{time}$ & $300s$\\
 \hline
 Warmup Time & $W$ & $60s$\\
 \hline
 Timestep Real Time & $t_{length}$ & $1s$\\
 \hline
 Recompute Std. Constants Freq. & $RSC$ & $100000$\\
 \hline
 Observation Stack Length & $k$ & $5$\\
 \hline
 Total Training Steps & - & $260000$\\
 \hline
 Number of AutoScale Actions & $ASA$ & $130000$\\
 \hline
 Number of Classifier-mod Actions & $CA$ & $50000$\\
 \hline
 Learning Rate & - & $3e-5$\\
 \hline
 Max Gradient Norm & - & $40$\\
 \hline
 Discount Factor & $\gamma$ & $0.9$\\
 \hline
 Replay Size & - & $200000$\\
 \hline
 Batch Size & $B$ & $100$\\
 \hline
 Polyak Coefficient & - & $0.995$\\
 \hline
 Entropy Coefficient & - & Learned Value\\
 \hline
 Initial Entropy Coefficient & $\eta$ & $1$\\
 \hline
 Target Entropy & - & $-\frac{1}{|M| \cdot |A|}$\\
 \hline
 \end{tabular}
\label{tab:MShyperparameters}
\end{table}

\LinesNumbered
\begin{algorithm}[!ht]
    \KwData{A microservice-based cloud service application}
    \KwIn{Initial policy network parameters $\theta$, policy network $\pi$, initial value-network parameters $\phi_1$, $\phi_2$, and action-value network $Q$}
    \nonl\myinput{Replay buffer $D$ with max capacity $N$}
    \KwOut{Fully-trained parameters $\phi_1$, $\phi_2$, and $\theta$}
    
    \vspace{1em}
    
    Collect $ASA$ samples using AutoScale Policy\\
    Store samples in $D$\\
    Compute $\mu, \sigma$ using $D$\\
    
    \vspace{1em}
    
    $t = 0$\\
    \Do{not converged}{
        Select random user count, $U \in [U_{min}, U_{max}]$\\
        Allocate maximum core count to all microservices\\
        Initialize workload generation with $U$ users\\
        Warmup for $W$ seconds\\
        
        Observe state $s$\\
        Mark timestep start, $t_{start} = t_{now}$\\
        
        \vspace{1em}
        
        \For{i in range($e_{time}$)}{
            $t = t + 1$\\
            select action $a\sim\pi_{\theta}(\cdot | s)$; execute $a$\\
            \uIf{$t < CA$}{
                Augment $a$ using the Classifier Network\\
            }
            
            \vspace{1em}
            
            \uIf{$t$ \% $RSC$ = $0$}{
                Recompute $\mu, \sigma$ using $D$\\
            }
            
            \vspace{1em}
            
            \uIf{$(t_{now} - t_{start}) < t_{length}$}{
                sleep($t_{length} - (t_{now} - t_{start})$)\\
            }
            $t_{start} = t_{now}$\\
            
            \vspace{1em}
            
            Observe next state $s'$ and reward $r$\\
            $d = True$ if $i > e_{time}$ else $False$\\
            Append $(s, a, r, s', d)$ to $D$\\
            
            Update parameters $\theta$, $\phi_1$, and $\phi_2$ via Algorithm \ref{alg:SoftActorCritic}\\
            
            $s = s'$\\
        }
    }
\caption{Training SAC for Core Allocation}
\label{alg:Training}
\end{algorithm}

% \begin{figure*}[!ht]
% \centering
% \subfloat[Mean Core Allocation]{\includegraphics[width=0.475\textwidth]{figures/social-mean_cpu.png}%
% \label{fig:socialAllocation_mean}}
% \hfil
% \hfil
% \rulesep
% \subfloat[Max Core Allocation]{\includegraphics[width=0.475\textwidth]{figures/social-max_cpu.png}%
% \label{fig:socialAllocation_max}}
% \hfil
% \caption{Core allocation results on Social Media service.}
% \label{fig:socialAllocation}
% \end{figure*}

The training procedure is shown in Algorithm \ref{alg:Training} and the associated hyperparameters and their symbols are shown in Table \ref{tab:MShyperparameters}. Algorithm~\ref{alg:Training} takes as input one of the two benchmarks, randomly initialized policy and value networks, and an empty replay buffer. In lines 1-2 of Algorithm \ref{alg:Training}, the agent collects data utilizing the industry-standard AutoScale  algorithm and does not perform updates. To ensure the data is diverse, Reclaimer applies uniform random noise to the core allocations given by AutoScale. The recommended allocations are allowed to vary by $\pm 5\%$. Random data collection before updating is common in DRL algorithms that utilize a replay buffer. 

%The purpose is to avoid overfitting the first samples collected \cite{mnih-human-level-2015, haarnoja-soft-2018}. However, in this domain, random core allocations provide very little information for learning a good policy. Instead, Reclaimer utilizes the AutoScale algorithm to provide good performance during the initial data collection and bias the learned policy toward a known, good policy. 

\begin{figure*}[!ht]
\centering
\subfloat[Mean Core Allocation]{\includegraphics[width=\columnwidth]{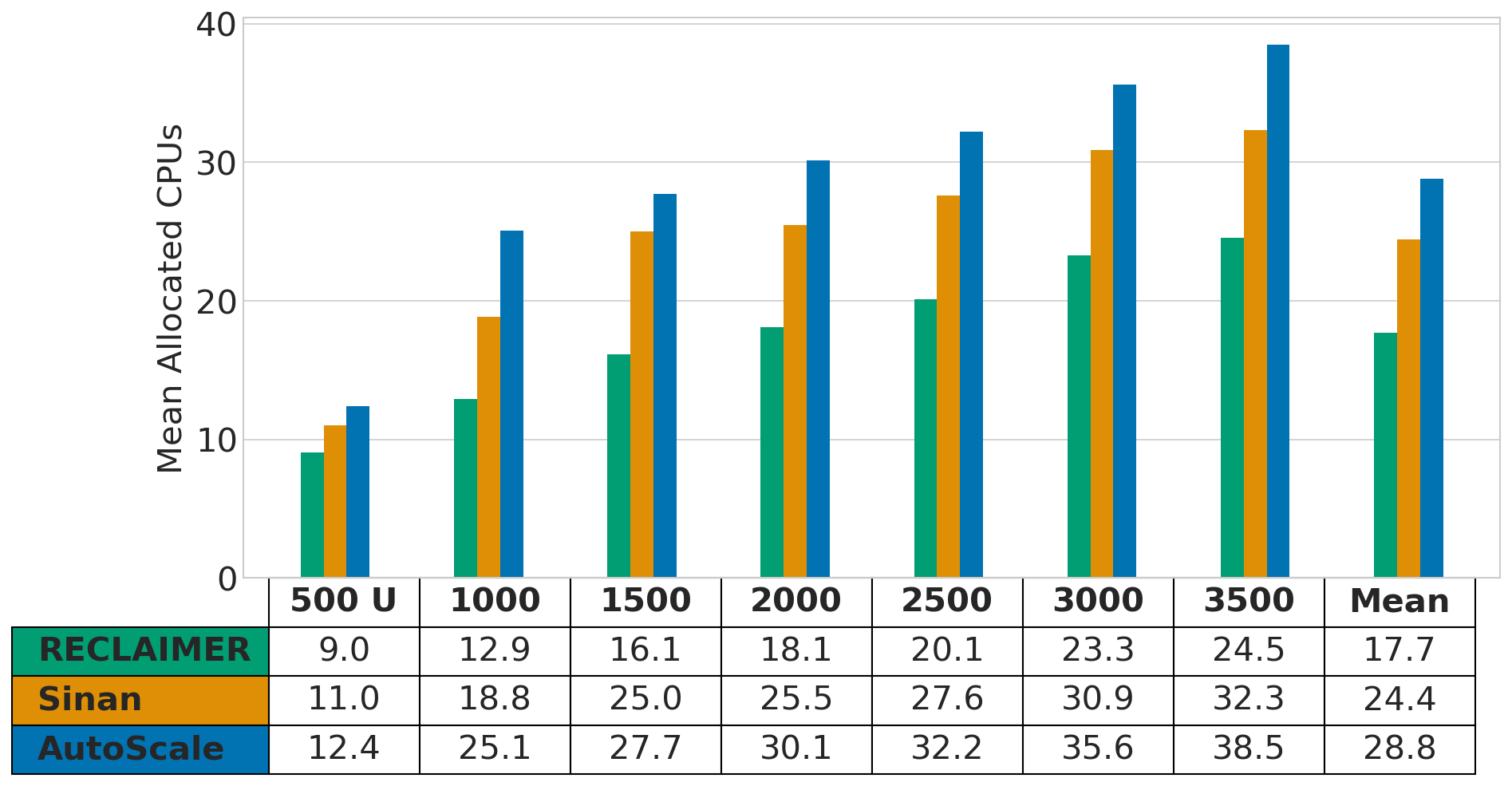}%
\label{fig:hotelAllocation_mean}}
\hfil
\hfil
\rulesep
\subfloat[Tail-latency]{\includegraphics[width=\columnwidth]{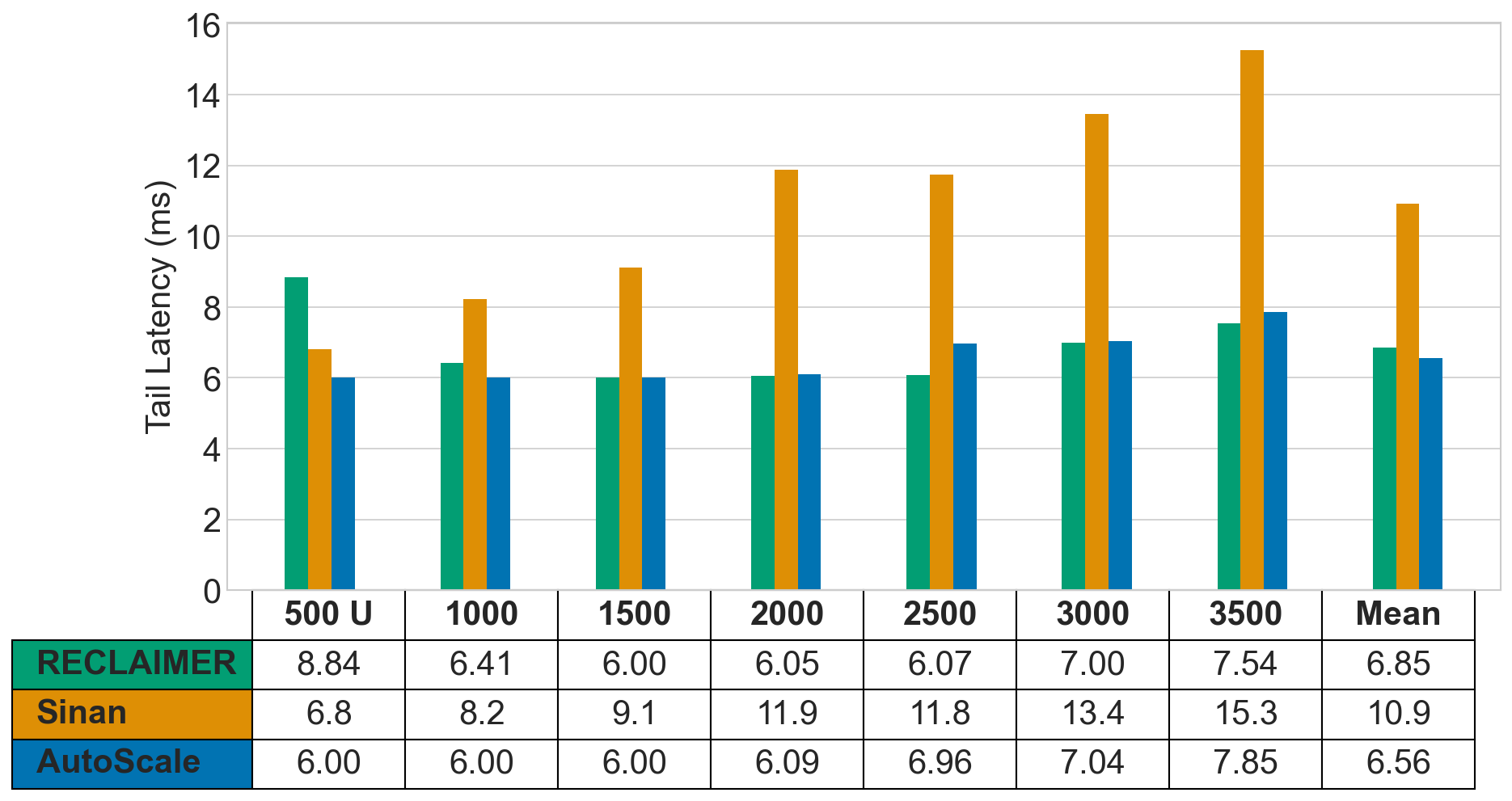}%
\label{fig:hotelLatency}}
\hfil
\label{fig:hotelAllocation}
\caption{Mean core allocation and Tail-latency results on Hotel Reservation service. The QoS requirement is 200ms.}
\end{figure*}

For the first Classifier-mod Actions ($CA$) iterations of the main training loop, lines 15-16 of the algorithm collect data by utilizing the network shown in Figure \ref{fig:InfoGain}. Given the action recommended by the RL agent, for each microservice, the classifier network considers $7$ possible actions at each timestep: scale the recommended CPU allocation down by $1/5/10\%$, keep the recommended core allocation, or scale the recommended core allocation up by $1/5/10\%$. The classifier network selects the candidate action which minimizes the allocated cores while having at least an estimated $80\%$ probability of meeting the QoS requirement. During the first $CA$ steps of the main loop, the probability of keeping the action proposed by the RL agent is linearly annealed from $0$ to $1$. Each element of the selected action vector is modified by $\pm 1\%$ using uniform random noise to inject more diversity in the data collected. The classifier network is retrained after each $e_{time}$ seconds using a maximum of $10000$ updates. In most situations, this network simply learns the prior probability of a QoS violation without considering the proposed action. The effect is such that the agent more frequently varies between meeting and missing the QoS requirement. Thus, the actions taken are on the boundary of `good' and `bad.'

Notably, in lines 6-9, it can be seen that a new workload is started and allowed to warm up for each randomly selected user count, $U$. This is opposed to the smooth transitions between the number of users that would occur in practice. This is due to limitations in the way workloads are started with Locust \cite{noauthor-locust-nodate}, nevertheless the method for starting workloads closely matches the code released for Sinan. For Social Media, Reclaimer is trained and tested using $20-200$ users, where each user generates a mean of $10$ requests per second according to an exponential distribution. The model for Hotel Reservation is trained and tested using $500-3500$ users, in which each user generates a mean of $1$ request per second according to an exponential distribution. The scripts for generating requests closely match the code released with Sinan to ensure a fair comparison. At line 3, the mean and standard deviation vectors are computed for the first time. Lines 5-28 compose the main training loop. Actions are selected using the policy and executed in line 14. Line 25 performs a single update per timestep. In line 18, the standardization constants are updated every $RSC$ timesteps, and line 20 allows the agent to idle for any leftover time each timestep. Lines 22-26 observe the outcome of the action, compute the done signal $d$ ($d=1$ if $t$ is the terminal timestep, and $d=0$ otherwise), store the full transition $(s, a, r, s', d)$ in the buffer, perform an update to the RL neural networks, and prepare the state $s$ for processing in the next timestep. Including the initial data collection using Autoscale and the classifier network, agents are trained for $260000$ timesteps, or $3$ days in wall-clock time when trained from scratch.

\subsection{Experiments on DeathStarBench}

All experiments are performed on a server with dual socket, Intel Xeon Gold 6230N 20-core, 40-thread CPUs, and two Tesla V100 GPUs. The operating system is Ubuntu 20.04.4 LTS. The CPU driver is set to acpi-cpufreq, and is set to use the `performance' governor with CPU frequency boosting enabled. These experiments are run on bare-metal. As discussed in Section \ref{sec:system_overview}, the CLI command \code{docker update --cpus=core\_allocation} is used to limit the available CPU cores for each microservice. The GPU driver is 510.54, with CUDA version 11.6. The Docker version is 20.10.13, and the docker-compose version is 1.29.2.

% \textcolor{blue}{The Reclaimer algorithm utilizes SAC \cite{haarnoja-soft-2018} to learn the RL policy, and a full description can be seen in Algorithm \ref{alg:Training}in Section \ref{sec:training}. For each application, fully trained models utilize $260000$ (3 days) training samples. With overhead due to workload generation and starting the applications, training and evaluation will take 4-5 days per application.}

% \begin{figure*}[!ht]
% \centering
% \subfloat[Mean Core Allocation]{\includegraphics[width=0.475\textwidth]{figures/hotel-mean_cpu.png}%
% \label{fig:hotelAllocation_mean}}
% \hfil
% \hfil
% \rulesep
% \subfloat[Max Core Allocation]{\includegraphics[width=0.475\textwidth]{figures/hotel-max_cpu.png}%
% \label{fig:hotelAllocation_max}}
% \hfil
% \caption{Core allocation results on Hotel Reservation.}
% \label{fig:hotelAllocation}
% \end{figure*}

We utilize two baselines in this work. Sinan \cite{zhang-sinan-2021} and Autoscale. For Sinan we used the open-source code provided by the authors to adapt it to work on a single server. Autoscale  is an algorithmic, industry standard method used by Amazon Web Services to dynamically scale resource allocations. The thresholds and scaling percentages for AutoScale were tuned to perform well on both benchmark applications, using values suggested by its authors. CPU allocation is increased by $10\%$ and $30\%$ when CPU utilization is between $[30\%, 50\%)$ and $[50\%, 100\%]$, respectively. This version of AutoScale reduces CPU allocation by $10\%$ when utilization is within $[0\%, 10\%]$.

Reclaimer was trained and evaluated on the same two applications from the DeathStarBench microservice benchmark suite \cite{gan-open-source-2019} that were originally used by Sinan  \cite{zhang-sinan-2021} in their evaluations:  Social Media and Hotel Reservation.
% Modified versions of the applications were released with the Sinan source code \cite{zhang-sinan-2021}, and were used to evaluate all algorithms. 
The scripts for generating workloads via Locust were identical to those released with Sinan \cite{zhang-sinan-2021}. The mean CPU core allocation, tail-latency, and max CPU core allocation are reported for each method. The primary objective of each algorithm is to meet the QoS requirement, while minimizing CPU allocation. Thus, lower core allocations are always desirable. Note that Reclaimer is not trained to minimize tail latency. Rather, Reclaimer is trained to minmize core allocatin while ensuring tail latency remains below the QoS requirement.

\begin{figure*}[!ht]
\centering
\subfloat[Mean Core Allocation]{\includegraphics[width=\columnwidth]{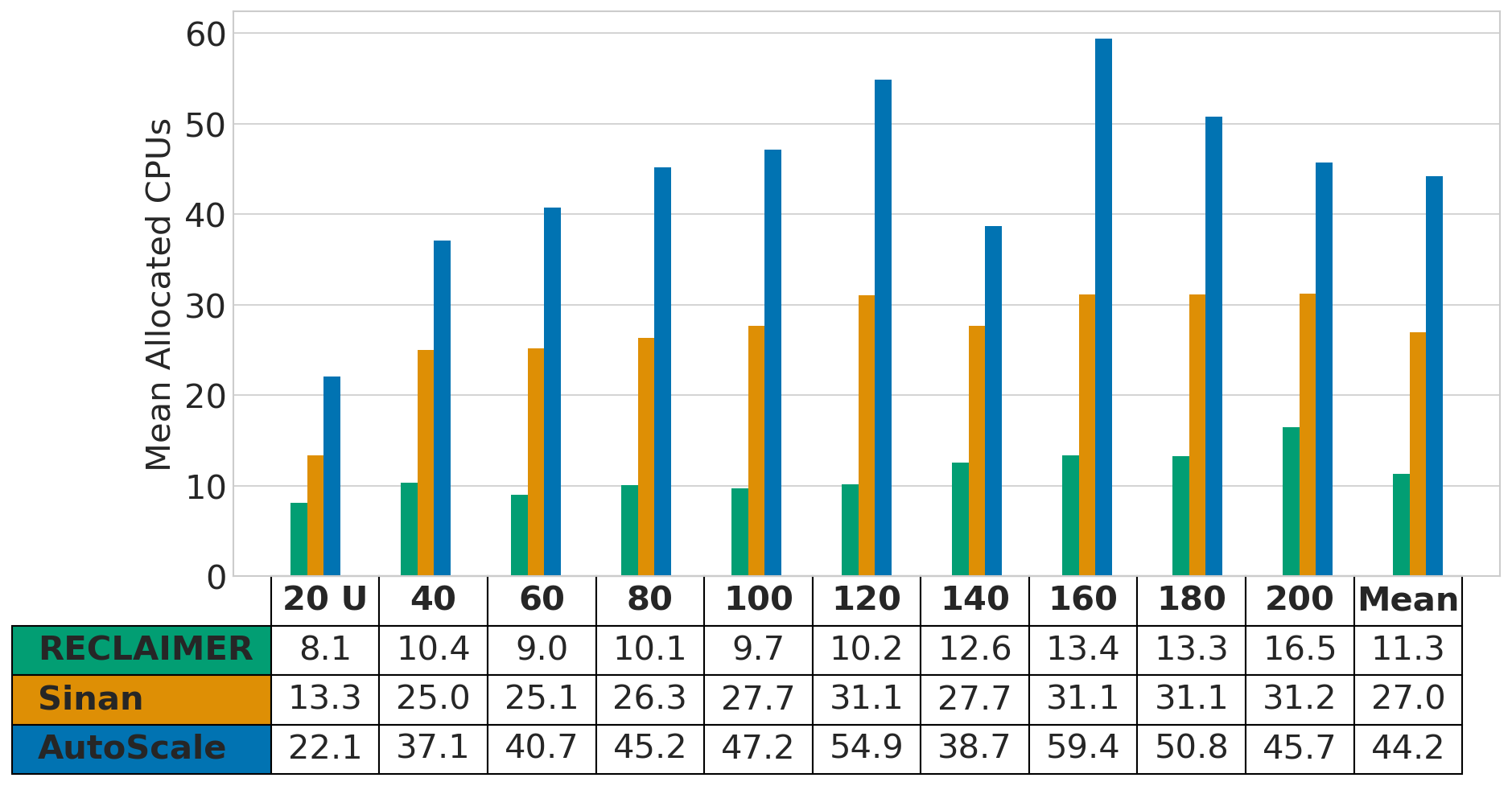}%
\label{fig:socialAllocation_mean}}
\hfil
\hfil
\rulesep
\subfloat[Tail-latency]{\includegraphics[width=\columnwidth]{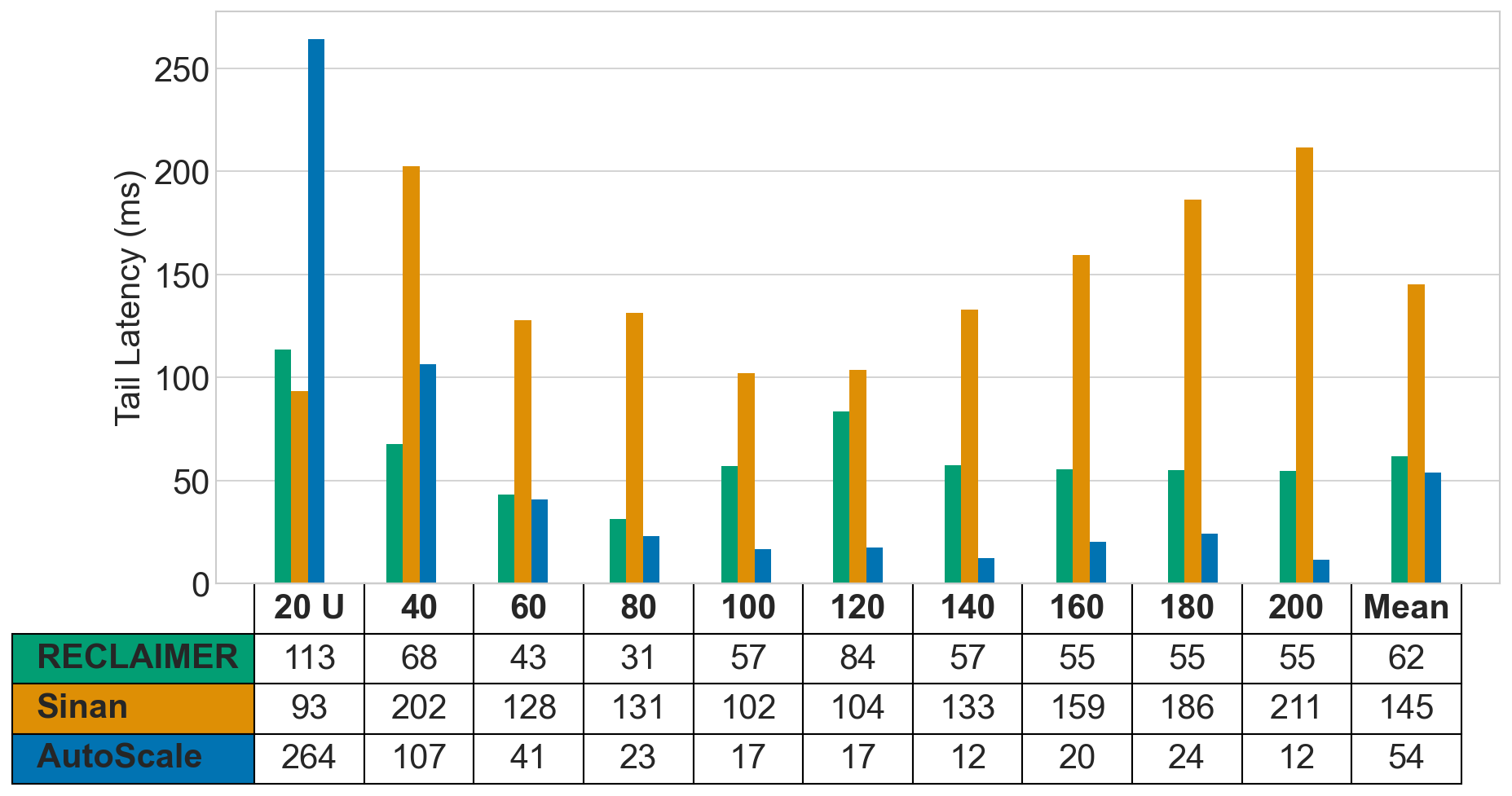}%
\label{fig:socialLatency}}
\hfil
\label{fig:socialAllocation}
\caption{Mean core allocation and Tail-latency results on Social Media service. The QoS requirement is 500ms.}
\end{figure*}

\section{Results and Discussion}
Core allocation results for Social Media are shown in Figure \ref{fig:socialAllocation_mean}. The y-axis shows the core allocations and the x-axis represents the user count. Each user count is evaluated for 5 minutes. Only Autoscale had a non-zero violation rate on Social Media, which was just $0.2\%$. Figure \ref{fig:socialAllocation_mean} shows the mean core allocation for each tested user count.
%, and Figure \ref{fig:socialAllocation_max} shows the maximum allocation. 
The rightmost entry in each graph is the mean across all user counts. Figure \ref{fig:socialLatency} shows the mean tail latency for each user count. 
%Figure \ref{fig:socialAllocation_mean} shows that 
Reclaimer reduces mean core allocation by $58.1\%$ relative to Sinan, and $74.4\%$ relative to AutoScale. Additionally, 
%Figure \ref{fig:socialAllocation_max} shows that 
Reclaimer increases the max core allocation by $9.7\%$ compared to Sinan, and decreases the max core allocation by $34.5\%$ compared to AutoScale. Finally, Figure \ref{fig:socialLatency} shows Reclaimer has a mean tail latency of $62$ms.

For Hotel Reservation, all methods had $0$ QoS violations. Figure \ref{fig:hotelAllocation_mean} indicates that Reclaimer reduced mean CPU allocation by $27.5\%$ relative to Sinan, and $38.4\%$ compared to AutoScale. Additionally, %Figure \ref{fig:hotelAllocation_max} shows that 
Reclaimer decreases the maximum core allocation by $28.0\%$ compared to Sinan on average, and reduces the maximum core allocation by $60.2\%$ compared to AutoScale. Finally, Reclaimer has a mean tail latency of $6.73$ms.
Three major metrics, mean CPU allocation, max CPU allocation, and violation rate indicate that Reclaimer is more effective at reducing the number of CPU cores allocated, while also meeting the QoS requirement $100\%$ of the time. When training Reclaimer from scratch on each benchmark application, Reclaimer outperforms both Autoscale and Sinan. 

\begin{figure*}[!tb]
\centering
\includegraphics[width=\textwidth]{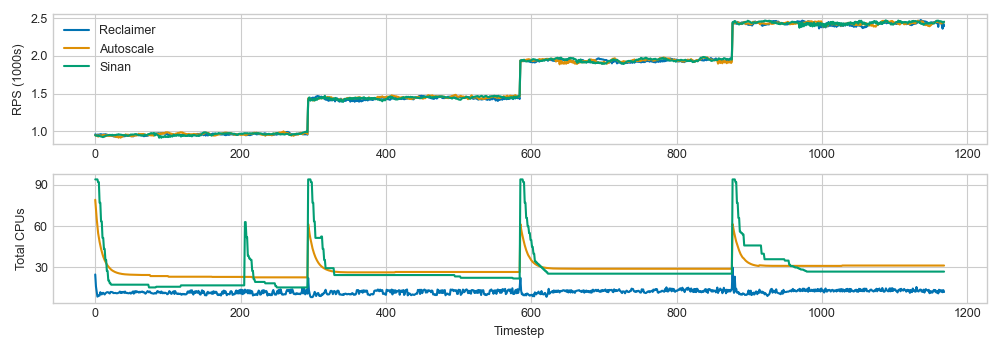}
\caption{Shown is per-timestep analysis for each method: Reclaimer, Sinan, and Autoscale. The application is Hotel Reservation Service. From top to bottom, each plot shows requests per second, and the total CPU core allocation to all microservices. The x-axis displays timesteps for both plots, where 1 timestep = 1 second.}
\label{fig:detailedPerformanceComparison}
\end{figure*}

Figure \ref{fig:detailedPerformanceComparison} displays a comparison of requests per second and total CPU allocation for each method throughout the execution of workloads. The top chart displays requests per second going up as the number of users is increased from 50 to 250. The bottom plot displays core allocation for each method. As expected, by dynamically responding to workloads, Reclaimer maintains the lowest core allocations consistently.

Notably, Reclaimer outperforms Sinan and Autoscale with respect to both tail-latency and mean CPU allocation. The latency and core counts reported in Figures \ref{fig:hotelAllocation_mean}, \ref{fig:hotelLatency}, \ref{fig:socialAllocation_mean}, and \ref{fig:socialLatency},  are mean values. A lower mean value for core counts is compatible with proactively allocating a sufficient number of cores to key microservices, eventually leading to lower core allocations at many later steps due to overall smaller queue depths. Figure \ref{fig:queuingEffect} indicates that sub-optimal core allocation decisions have delayed and long-term effects on the tail latency of applications. Neither Autoscale nor Sinan are pareto optimal with respect to tail latency and total core allocation, making it possible for other methods such as Reclaimer to discover policies that improve both metrics.

% Although this evaluation was performed on a single node, Sinan was originally evaluated in a multi-node cluster, which introduces network latency and congestion. While Sinan did not model network latency and congestion explicitly, their effect was captured implicitly through the values computed for the feature vectors used in its ML models. Because Reclaimer's features are a superset of Sinan's features, we expect that Reclaimer has sufficient information to respond to these network-related effects, and the performance improvement brought by Reclaimer to transfer to multi-node settings. Additionally, in both this work and the original Sinan \cite{zhang-sinan-2021}, the workload was never large enough to cause a significant number of QoS violations. Because Reclaimer is more effective than Sinan at all tested workloads, we expect it to remain so when the load becomes large enough to cause QoS violations. However, exploring the performance of Reclaimer and Sinan in a multi-node cluster with more demanding workloads remains a useful direction for future work.

% \begin{figure}[!tb]
% \centering
% \includegraphics[width=\columnwidth]{figures/hotel-mean_cpu.png}
% \caption{Core allocation results on Hotel Reservation.}
% \label{fig:hotelAllocation_mean}
% \end{figure}

\subsection{Online Learning}
Reclaimer collects its initial training data from the Autoscale policy, ensuring that, in the worst case, it can match the performance of a known, good policy for the application at hand. Additionally, this enables future human intervention to improve Reclaimer. As human designers develop new policies for resource management, Reclaimer can collect data and improve upon those policies. To include data from a newly engineered policy, a designer would replace the data collection from Autoscale in line 1 of Algorithm \ref{alg:Training} with the new data collection policy, or collect data with both Autoscale and the new engineered policy in a separate step. After this initial phase of safe data collection, the next phase of data collection, shown in lines 15-16 of Algorithm \ref{alg:Training} allows Reclaimer to collect data on the boundary of good and bad decisions, and improves upon the solution given by Autoscale. Finally, after all data collection, Reclaimer is deployed and allowed to learn online, consistently improving performance in lines 12-28 of Algorithm \ref{alg:Training}. In our experiments, $e_{time}$ is $3$ days. However, in a real deployment, the loop in lines 12-28 can run ad infinitum, allowing Reclaimer to continue collecting data throughout an application's deployment and use it to adjust rapidly to any changes in the workload and microservices.

% no need for tuning per-application
\subsection{Generality of Reclaimer}
Autoscale used the same parameters for both applications in these experiments. On the other hand, Sinan's hyperparameters are tuned for each application individually. As shown in Table \ref{tab:MShyperparameters}, 
Reclaimer uses a single set of hyperparameters to learn an effective policy for both applications. In practice, this significantly reduces the complexity of deploying Reclaimer for any given microservice-based service. A fixed set of hyperparameters is important for making Reclaimer accessible to microservice application developers, so that sophisticated ML expertise is not required to ensure core allocations are scaled effectively after each update to the application.

% Transfer experiment
\subsection{Deployment on Different Applications through Transfer Learning}
\label{sec:transferLearning}

% \begin{figure*}[!ht]
% \centering
% \subfloat[]{\includegraphics[width=0.475\textwidth]{figures/transfer-mean_cpu.png}%
% \label{fig:MSTransfer_mean}}
% \hfil
% \hfil
% \rulesep
% \subfloat[]{\includegraphics[width=0.475\textwidth]{figures/transfer-tail_latency.png}%
% \label{fig:MSTransfer_max}}
% \hfil
% \caption{Pictured is Reclaimer trained with limited data in a transfer learning setting. `Transfer Learning' is Reclaimer after being trained first on Hotel Reservation then transferred and fine-tuned on Social Media using limited data, and `Train from Scratch' is Reclaimer trained with a random initialization on a limited dataset collected from Social Media. \textbf{(a)} shows mean core allocations, while \textbf{(b)} shows tail latency.}
% \label{fig:MSTransfer}
% \end{figure*}

\begin{figure}[!htb]
\centering
\includegraphics[width=\columnwidth]{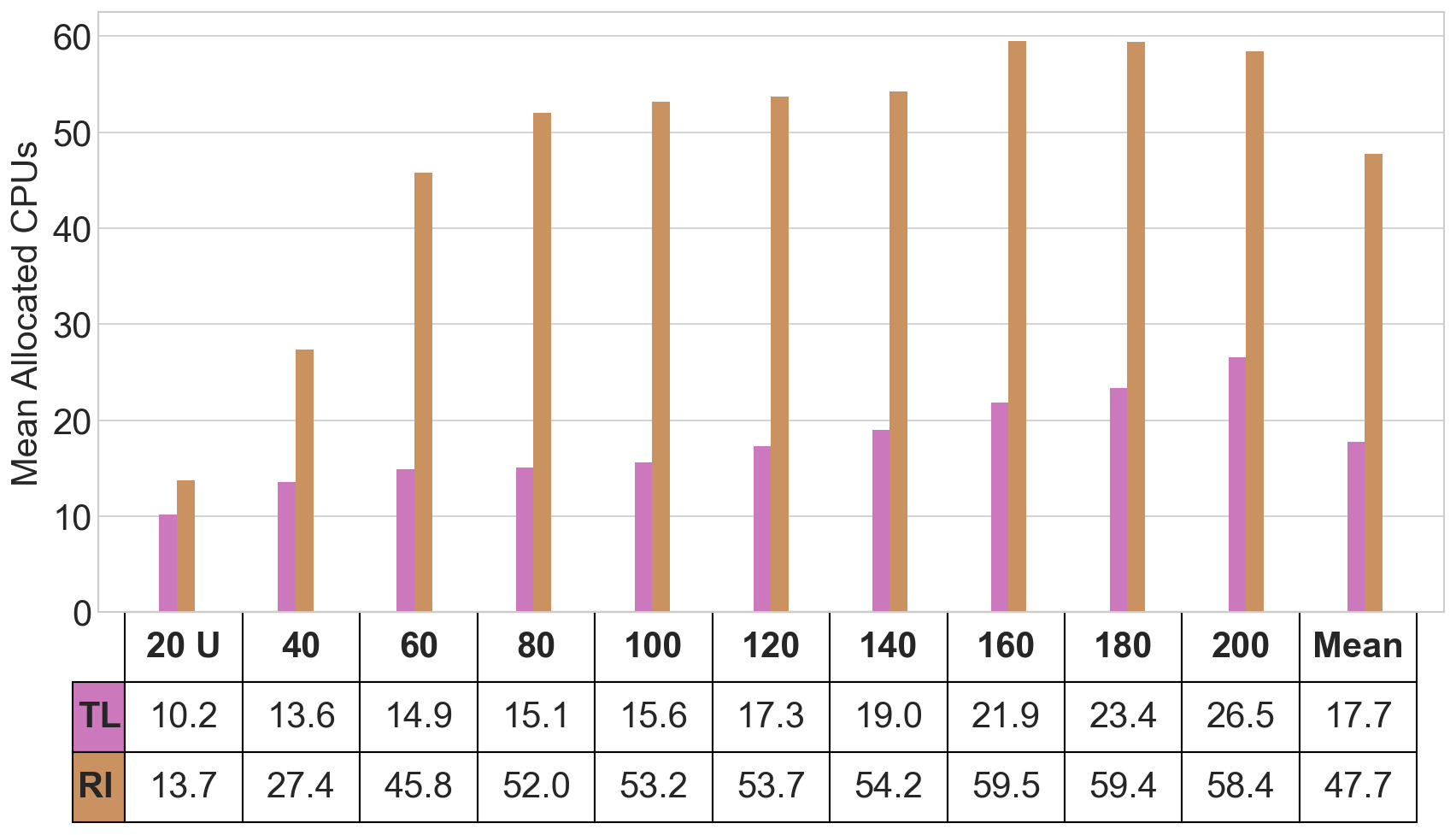}
\caption{Pictured are mean core allocation results after Reclaimer is trained with limited data in a transfer learning setting. `TL' (Transfer Learning)  is Reclaimer after being trained first on Hotel Reservation then transferred and fine-tuned on Social Media using limited data, and `RI' (Random Initialization) is Reclaimer trained with a random initialization on a limited dataset collected from Social Media.}
\label{fig:MSTransfer}
\end{figure}

Recall that Reclaimer re-uses $~99.95\%$ of model parameters to process each individual microservice. This enables Reclaimer to rapidly adapt to new microservices and changes to the entire microservice dependency graph (which is neither explicitly represented nor given as input to the model). To demonstrate the ability of Reclaimer to adapt to significant changes in the microservice dependency graph, we developed a worst-case-scenario, where the agent was trained on Hotel Reservation, then deployed on Social Media. In this situation, nearly all aspects of the underlying problem are made non-stationary. Every microservice in the overall service has changed, the dependencies among these new microservices are different, the user counts are different, the QoS requirement has changed, and the number of CPU cores available for each microservice has changed. To facilitate this change, Reclaimer collects $70000$ timesteps of data from Social Media using the AutoScale policy, and $50000$ timesteps of data utilizing the classification network shown in Figure \ref{fig:InfoGain}. This is $46.1\%$ of the data needed to train Reclaimer with a random initialization. Core allocation results are shown in Figure \ref{fig:MSTransfer}. After performing $120000$ updates, a randomly initialized model trained on Social Media has a $38\%$ violation rate, and allocates $322\%$ more cores than the fully-trained model on average. On the other hand, the transfer learning model which is trained first on Hotel Reservation then fine-tuned on $120000$ timesteps of data from Social Media, allocates just $56.6\%$ more cores than the fully trained model for Social Media, with no QoS violations. 

Large-scale, real-world traces from microservice applications would exhibit forms of non-stationarity, such as gradual shifts in the number of user requests and changes in user behavior, but these do not yet exist for our microservice benchmark applications. While this experiment does not test all types of non-stationarity, it demonstrates two major strengths of Reclaimer. First, the policy it learns generalizes to completely new, unseen microservices. 
% This is expected because Reclaimer uses the same set of parameters to process every microservice in the dependency graph, forcing the model to learn general patterns that apply to classes of microservices. 
Second, it demonstrates that Reclaimer can reduce the sample complexity by adapting a previously trained policy, even when the new application is very different from the previous one.

\begin{figure*}[!ht]
\centering
\subfloat[Nginx Web Server]{\includegraphics[width=0.35\textwidth]{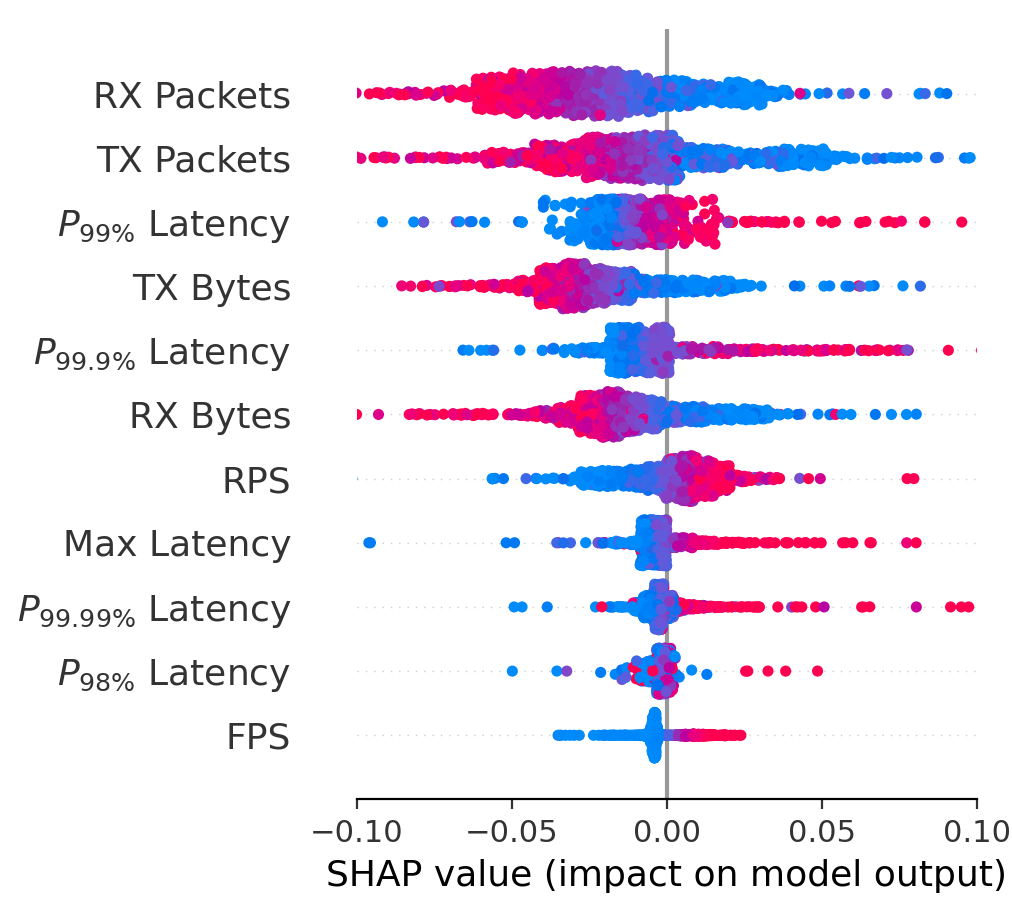}%
\label{fig:SHAP_nginx}}
\hfil
\hfil
\subfloat[Media Filter]{\includegraphics[width=0.41\textwidth]{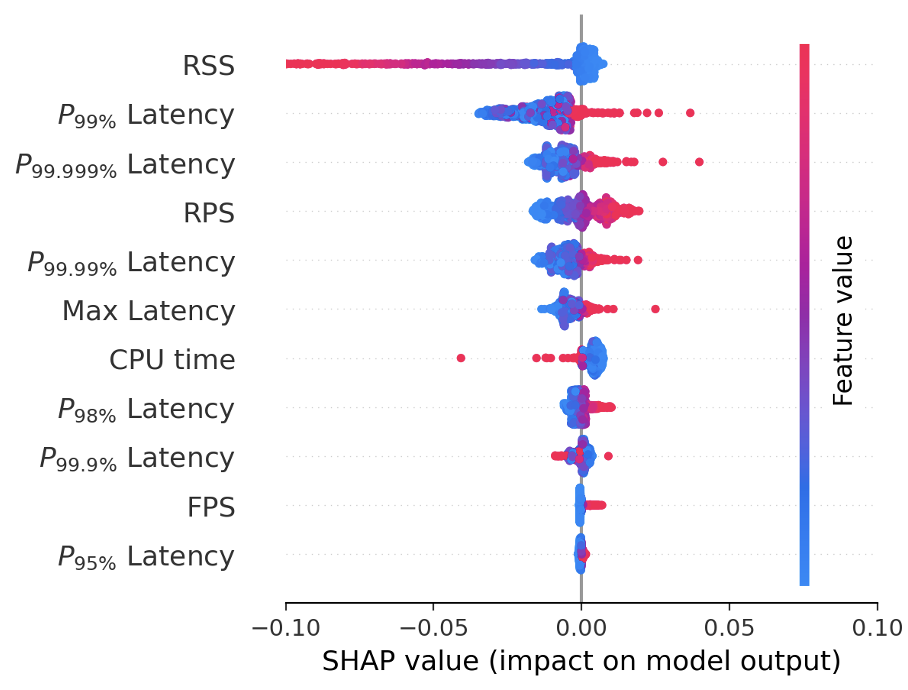}%
\label{fig:SHAP_mediafilter}}
\hfil
\caption{SHAP values, a measure of feature importance, are computed using $1000$ randomly sampled inputs for the policy network. The vertical axis shows the feature names in descending order according to the mean absolute SHAP value for all sampled points. The horizontal position of each point represents the SHAP value for the corresponding feature. The color of each point represents the relative input value of the feature, where red indicates the input feature had a large value, purple represents a mean value, and blue represents a low value. Both \textbf{(a)} and \textbf{(b)} represent microservices from Social Media, and the SHAP values for each were computed using a single, fully-trained policy. Outliers were omitted such that all displayed points have SHAP values in the range $[-0.1, 0.1]$. Only the top $11$ features are shown for each microservice.}
\label{fig:SHAP_Values}
\end{figure*}

\subsection{Introspection using SHAP}

To understand the decisions made by Reclaimer, we utilize SHAP \cite{lundberg-unified-2017}, a game theoretic approach designed to help interpret the output of arbitrary ML models. SHAP values measure the relative impact of a feature on the model's output, e.g. a SHAP value of $0.1$ indicates that the model's output increased by $0.1$ for a particular feature value. Figure \ref{fig:SHAP_Values} shows the SHAP values for two distinct microservices from Social Media: Nginx Web Server and Media Filter. All SHAP values were produced using the policy network shown in Figure \ref{fig:MSPolicyNetwork}. Important features that are the most intuitive are those relating to the current latency distribution, $P_{x\%}$. For both microservices, high values (red dots) for any percentile of the end-to-end latency yield correspondingly large, positive SHAP values, and vice-versa (i.e. high latency causes the model to allocate more cores). %Conversely, low values (blue dots) for any latency percentile causes the model to decrease the core allocations. Because Reclaimer is optimizing directly for $99^{th}$ percentile latency, it is expected that $P_{99\%}$ is the most important latency feature for both microservices.

Nginx Web Server is a HTTP server and reverse proxy, and serves as the first microservice that all user requests interact with. Figure \ref{fig:SHAP_nginx} shows that the most important features for making core allocation decisions for this microservice include received packets, received bytes, transmitted packets, and transmitted bytes. Interestingly, as these values become \textit{larger} than average, the policy allocates \textit{fewer} cores. These features do not accumulate while requests are waiting on microservice queues, and only increase after the microservice has interacted with incoming requests or sent outgoing requests. Thus, when the values of these features are \textit{too} large, it indicates the microservice is working faster than needed under the QoS constraints, hence its core allocation could be lowered while still maintaining acceptable latency.

The client component of Media Filter receives posts with attached images, and uses a trained model to determine if those images contain inappropriate content, an operation that is CPU intensive. Afterward, the images are uploaded to the server component of Media Filter, which forwards the posts to the appropriate down-stream microservices. Because the server component only reads images from clients and forwards them, it is significantly less CPU intensive. Figure \ref{fig:SHAP_mediafilter} shows that as Requests Per Second (RPS) increases and more images need processed by the ML model, the RL policy increases the core allocation. As Resident Set Size (RSS) increases because many images have already been processed by the model and are waiting in memory to be forwarded to down-stream microservices, the RL policy decreases the core allocation.

The significantly different ranking of the features between Nginx Web Server and Media Filter show how a single trained policy can have a very different, nuanced view of the behaviors of distinct microservices.

\subsection{Limitations}
Although the methods are evaluated on a single node due to limited server infrastructure availability, we expect the performance improvements brought by Reclaimer to transfer to multi-node settings. Because Reclaimer's features are a superset of Sinan's features, we expect that Reclaimer has sufficient information to respond to network-related effects. Additionally, in both this work and the original Sinan \cite{zhang-sinan-2021}, the workload was never large enough to cause a significant number of QoS violations. Because Reclaimer is more effective than Sinan at all tested workloads, we expect it to remain so when the load becomes large enough to cause QoS violations. However, exploring the performance of Reclaimer and Sinan in a multi-node cluster with more demanding workloads remains a useful direction for future work.

\section{Related Work}

% microservices behavior
DeathStarBench \cite{gan-open-source-2019} introduced an open-source microservice benchmark suite with six microservice-based benchmarks to evaluate the implications of the microservices model. SEER \cite{gan-seer-2018} and SAGE \cite{gan-sage-2021} use machine learning methods to debug performance issues with microservice-based cloud services, and prescribe solutions to those performance issues. In \cite{guo-who-2019}, real application traces from Alibaba datacenters were collected over an 8-day period to determine the resource efficiency of various types of services.

% minimizing tail latency
Many approaches seek to minimize the latency of requests independent of resource considerations, such as \cite{hou-alphar-2021, lin-ant-2019}.
There is a broad focus on co-locating latency-critical monolithic services with background tasks or other services such that the co-located applications do not interfere with the same resources \cite{yang-bubble-flux-2013, patel-clite-2020, chen-parties-2019, yang-miras-2019, garefalakis-medea-2018, hindman-mesos-2011}. Other approaches utilize Deep RL to schedule distributed data stream processing workers to machines such that communication delays are avoided and end-to-end latency is reduced \cite{xu-experience-driven-2018, li-model-free-2018}. Others have applied DRL based techniques for large-scale resource allocation and service provisioning in Internet-of-Things (IoT) \cite{IoTDRL}.
Pensieve \cite{mao-neural-2017} utilizes DeepRL to proactively select bitrates for video streaming chunks such that streaming performance is maximized and user experience is improved. DeepRecSys \cite{gupta-deeprecsys-2020} specifically targets recommendation systems. By considering characteristics of input queries, their arrival rates, the architecture of the recommender system, and the underlying hardware, DeepRecSys is able to optimize the batch size of queries and leverage parallelism to increase throughput in a variety of recommendation systems. Finally, 
GRAF \cite{park-graf-2021} and pHPA \cite{choi-phpa-2021} are two recent autoscaling approaches which proactively scale resources via machine learning methods to minimize latency while meeting service level objectives.

% minimizing resource consumption in monoliths
% \cite{noauthor-aws-nodate}
Autoscale  is an industry-standard approach to scaling resources proportional to workloads. AutoScale computes a ratio of resource utilization to resource allocation, then scales the resource allocation up/down when the ratio reaches manually tuned thresholds. 
%This approach can be tuned such that the QoS constraints of a latency-critical service can be met while also reducing overall resource allocation. 
Other methods utilize approximate computing \cite{kulkarni-leveraging-2018} and DVFS \cite{yang-powerchief-2017} to meet QoS requirements, while saving resources. Moreover, in HyScale \cite{HyScale} and CoScal \cite{CoScal} approaches, QoS constraints have been met by reactive algorithms by effectively combining horizontal and vertical scaling approaches. However, Reclaimer adapts to runtime changes to in the number and behavior of microservices.
Additionally, a large body of work is concerned with co-locating latency-critical monolithic applications with throughput-oriented background job, such as \cite{li-dynamic-2020, nishtala-hipster-2017}. 
These approaches aim to identify background tasks which do not interfere with resources needed by latency-critical tasks. 
\citet{thamsen-learning-2017} seeks to improve resource utilization in data-parallel, throughput oriented processing jobs by utilizing RL to decide which applications should be co-located within the cluster.

FIRM \cite{qiu-firm-2020} {\it reactively} adjusts resources for microservices in response to QoS violations, using SVMs to detect the microservices causing QoS violations and RL to recover. Every microservice is specially instrumented to extract and store tracing information in a centralized database. A Support Vector Machine is trained to identify microservices which have already caused QoS violations, and DRL is used to recover from QoS violations, while also minimizing resource utilization. Unlike our proactive approach, FIRM is reactive to QoS violations, i.e. it will adjust allocations only after a QoS violation was detected. 
Sinan \cite{zhang-sinan-2021} first predicts future QoS violations, then makes resource adjustments such that QoS violations are avoided and CPU utilization is minimized, using a combination of trained models and hand-engineered policies. Their approach utilizes a data collection step in which a bandit algorithm collects data to train a deep neural network, and the output of the deep neural network is used as input to a gradient boosted decision tree. The decision tree predicts the probability that a QoS violation occurs in the future. Using probabilities from the decision tree, core allocations are scaled according to a hand-engineered policy to avoid any potential QoS violations. In \cite{QoSAware} a resource manager determines the contention due to communication between different nodes and maps microservices from busy to idle nodes.
In contrast to these approaches, our proposed model does not require per-application hyperparameter tuning, can adapt to a variable number of microservices, is adaptive to changes in individual microservices, does not require explicit engineering of the inputs for use in convolutional layers, and is dynamic to changes in the available resources without gathering new data and retraining. Additionally, our DRL approach enables the distribution of the training data to match the distribution of the evaluation data as the policy changes.

\section{Conclusions}

We propose Reclaimer, a DRL-based system for core allocations
%which matches the dynamic nature of the microservice model. Our proposed method learns online to 
that adapts to changes in the workload,  microservices, and underlying hardware. Reclaimer preemptively adjusts core allocations to meet QoS requirements 100\% of the time, while allocating on average $38.4\% - 74.4\%$ fewer CPU cores than the industry standard approach Autoscale, and $27.5\% - 58.1\%$ fewer cores than Sinan \cite{zhang-sinan-2021}. We demonstrate that Reclaimer is adaptive enough to handle extreme changes in the microservices. Pretraining on a different microservice-based application was shown to speed up the convergence of Reclaimer by approximately $2\times$ when compared to a model trained with a random parameter initialization. 
Model introspection shows that the policy has the capacity to learn that feature importance varies between distinct microservices when optimizing core allocations. Altogether, Reclaimer is a highly adaptive, proactive approach to resource allocation for cloud microservices that outperforms prior approaches without the need for per-application tuning.

\bibliographystyle{ACM-Reference-Format}
\bibliography{hpdc23}


\begin{thebibliography}{49}

%%% the Web address.

\ifx \showCODEN    \undefined \def \showCODEN     #1{\unskip}     \fi
\ifx \showDOI      \undefined \def \showDOI       #1{#1}\fi
\ifx \showISBNx    \undefined \def \showISBNx     #1{\unskip}     \fi
\ifx \showISBNxiii \undefined \def \showISBNxiii  #1{\unskip}     \fi
\ifx \showISSN     \undefined \def \showISSN      #1{\unskip}     \fi
\ifx \showLCCN     \undefined \def \showLCCN      #1{\unskip}     \fi
\ifx \shownote     \undefined \def \shownote      #1{#1}          \fi
\ifx \showarticletitle \undefined \def \showarticletitle #1{#1}   \fi
\ifx \showURL      \undefined \def \showURL       {\relax}        \fi
% invisible to TeX
\providecommand\bibfield[2]{#2}
\providecommand\bibinfo[2]{#2}
\providecommand\natexlab[1]{#1}
\providecommand\showeprint[2][]{arXiv:#2}

\bibitem[noa(2021a)]%
        {noauthor-cfs-nodate}
 \bibinfo{year}{2021}\natexlab{a}.
\newblock \bibinfo{title}{{CFS} {Scheduler} — {The} {Linux} {Kernel}
  documentation}.
\newblock
\newblock
\urldef\tempurl%
\url{https://www.kernel.org/doc/html/latest/scheduler/sched-design-CFS.html}
\showURL{%
\tempurl}


\bibitem[noa(2021b)]%
        {noauthor-locust-nodate}
 \bibinfo{year}{2021}\natexlab{b}.
\newblock \bibinfo{title}{Locust - {A} modern load testing framework}.
\newblock
\newblock
\urldef\tempurl%
\url{https://locust.io/}
\showURL{%
\tempurl}


\bibitem[noa(2021c)]%
        {noauthor-runtime-2021}
 \bibinfo{year}{2021}\natexlab{c}.
\newblock \bibinfo{title}{Runtime options with {Memory}, {CPUs}, and {GPUs}}.
\newblock
\newblock
\urldef\tempurl%
\url{https://docs.docker.com/config/containers/resource-constraints/}
\showURL{%
\tempurl}


\bibitem[Achiam(2018)]%
        {achiam-spinning-2018}
\bibfield{author}{\bibinfo{person}{Joshua Achiam}.}
  \bibinfo{year}{2018}\natexlab{}.
\newblock \showarticletitle{Spinning {Up} in {Deep} {Reinforcement}
  {Learning}}.
\newblock  (\bibinfo{year}{2018}).
\newblock


\bibitem[Bellemare et~al\mbox{.}(2013)]%
        {bellemare-arcade-2013}
\bibfield{author}{\bibinfo{person}{M.~G. Bellemare}, \bibinfo{person}{Y.
  Naddaf}, \bibinfo{person}{J. Veness}, {and} \bibinfo{person}{M. Bowling}.}
  \bibinfo{year}{2013}\natexlab{}.
\newblock \showarticletitle{The {Arcade} {Learning} {Environment}: {An}
  {Evaluation} {Platform} for {General} {Agents}}.
\newblock \bibinfo{journal}{\emph{Journal of Artificial Intelligence Research}}
   \bibinfo{volume}{47} (\bibinfo{date}{June} \bibinfo{year}{2013}),
  \bibinfo{pages}{253--279}.
\newblock


\bibitem[Chen et~al\mbox{.}(2019)]%
        {chen-parties-2019}
\bibfield{author}{\bibinfo{person}{Shuang Chen}, \bibinfo{person}{Christina
  Delimitrou}, {and} \bibinfo{person}{José~F. Martínez}.}
  \bibinfo{year}{2019}\natexlab{}.
\newblock \showarticletitle{{PARTIES}: {QoS}-{Aware} {Resource} {Partitioning}
  for {Multiple} {Interactive} {Services}}. In
  \bibinfo{booktitle}{\emph{Proceedings of the {Twenty}-{Fourth}
  {International} {Conference} on {Architectural} {Support} for {Programming}
  {Languages} and {Operating} {Systems}}} \emph{(\bibinfo{series}{{ASPLOS}
  '19})}. \bibinfo{publisher}{Association for Computing Machinery},
  \bibinfo{address}{New York, NY, USA}, \bibinfo{pages}{107--120}.
\newblock
\showISBNx{978-1-4503-6240-5}
\urldef\tempurl%
\url{https://doi.org/10.1145/3297858.3304005}
\showDOI{\tempurl}


\bibitem[Choi et~al\mbox{.}(2021)]%
        {choi-phpa-2021}
\bibfield{author}{\bibinfo{person}{Byungkwon Choi}, \bibinfo{person}{Jinwoo
  Park}, \bibinfo{person}{Chunghan Lee}, {and} \bibinfo{person}{Dongsu Han}.}
  \bibinfo{year}{2021}\natexlab{}.
\newblock \showarticletitle{{pHPA}: {A} {Proactive} {Autoscaling} {Framework}
  for {Microservice} {Chain}}. In \bibinfo{booktitle}{\emph{5th
  {Asia}-{Pacific} {Workshop} on {Networking} ({APNet} 2021)}}
  \emph{(\bibinfo{series}{{APNet} 2021})}. \bibinfo{publisher}{Association for
  Computing Machinery}, \bibinfo{address}{New York, NY, USA},
  \bibinfo{pages}{65--71}.
\newblock
\showISBNx{978-1-4503-8587-9}
\urldef\tempurl%
\url{https://doi.org/10.1145/3469393.3469401}
\showDOI{\tempurl}


\bibitem[Coumans and Bai(2016)]%
        {coumans-pybullet-2016}
\bibfield{author}{\bibinfo{person}{Erwin Coumans} {and} \bibinfo{person}{Yunfei
  Bai}.} \bibinfo{year}{2016}\natexlab{}.
\newblock \bibinfo{title}{{PyBullet}, a {Python} module for physics simulation
  for games, robotics and machine learning}.
\newblock
\newblock
\urldef\tempurl%
\url{http://pybullet.org}
\showURL{%
\tempurl}


\bibitem[Esposito et~al\mbox{.}(2016)]%
        {esposito-challenges-2016}
\bibfield{author}{\bibinfo{person}{Christian Esposito},
  \bibinfo{person}{Aniello Castiglione}, {and}
  \bibinfo{person}{Kim-Kwang~Raymond Choo}.} \bibinfo{year}{2016}\natexlab{}.
\newblock \showarticletitle{Challenges in {Delivering} {Software} in the
  {Cloud} as {Microservices}}.
\newblock \bibinfo{journal}{\emph{IEEE Cloud Computing}} \bibinfo{volume}{3},
  \bibinfo{number}{5} (\bibinfo{date}{Sept.} \bibinfo{year}{2016}),
  \bibinfo{pages}{10--14}.
\newblock
\showISSN{2325-6095}
\urldef\tempurl%
\url{https://doi.org/10.1109/MCC.2016.105}
\showDOI{\tempurl}
\newblock
\shownote{Conference Name: IEEE Cloud Computing}.


\bibitem[Fu et~al\mbox{.}(2021)]%
        {QoSAware}
\bibfield{author}{\bibinfo{person}{Kaihua Fu}, \bibinfo{person}{Wei Zhang},
  \bibinfo{person}{Quan Chen}, \bibinfo{person}{Deze Zeng},
  \bibinfo{person}{Xin Peng}, \bibinfo{person}{Wenli Zheng}, {and}
  \bibinfo{person}{Minyi Guo}.} \bibinfo{year}{2021}\natexlab{}.
\newblock \showarticletitle{QoS-Aware and Resource Efficient Microservice
  Deployment in Cloud-Edge Continuum}. In \bibinfo{booktitle}{\emph{2021 IEEE
  International Parallel and Distributed Processing Symposium (IPDPS)}}.
  \bibinfo{pages}{932--941}.
\newblock
\urldef\tempurl%
\url{https://doi.org/10.1109/IPDPS49936.2021.00102}
\showDOI{\tempurl}


\bibitem[Gan et~al\mbox{.}(2021)]%
        {gan-sage-2021}
\bibfield{author}{\bibinfo{person}{Yu Gan}, \bibinfo{person}{Mingyu Liang},
  \bibinfo{person}{Sundar Dev}, \bibinfo{person}{David Lo}, {and}
  \bibinfo{person}{Christina Delimitrou}.} \bibinfo{year}{2021}\natexlab{}.
\newblock \showarticletitle{Sage: practical and scalable {ML}-driven
  performance debugging in microservices}. In
  \bibinfo{booktitle}{\emph{Proceedings of the 26th {ACM} {International}
  {Conference} on {Architectural} {Support} for {Programming} {Languages} and
  {Operating} {Systems}}} \emph{(\bibinfo{series}{{ASPLOS} 2021})}.
  \bibinfo{publisher}{Association for Computing Machinery},
  \bibinfo{address}{New York, NY, USA}, \bibinfo{pages}{135--151}.
\newblock
\showISBNx{978-1-4503-8317-2}
\urldef\tempurl%
\url{https://doi.org/10.1145/3445814.3446700}
\showDOI{\tempurl}


\bibitem[Gan et~al\mbox{.}(2018)]%
        {gan-seer-2018}
\bibfield{author}{\bibinfo{person}{Yu Gan}, \bibinfo{person}{Meghna Pancholi},
  \bibinfo{person}{Siyuan Hu}, \bibinfo{person}{Dailun Cheng},
  \bibinfo{person}{Yuan He}, {and} \bibinfo{person}{Christina Delimitrou}.}
  \bibinfo{year}{2018}\natexlab{}.
\newblock \showarticletitle{Seer: {Leveraging} {Big} {Data} to {Navigate} the
  {Increasing} {Complexity} of {Cloud} {Debugging}}.
\newblock
\urldef\tempurl%
\url{https://www.usenix.org/conference/hotcloud18/presentation/gan}
\showURL{%
\tempurl}


\bibitem[Gan et~al\mbox{.}(2019a)]%
        {gan-open-source-2019}
\bibfield{author}{\bibinfo{person}{Yu Gan}, \bibinfo{person}{Yanqi Zhang},
  \bibinfo{person}{Dailun Cheng}, \bibinfo{person}{Ankitha Shetty},
  \bibinfo{person}{Priyal Rathi}, \bibinfo{person}{Nayan Katarki},
  \bibinfo{person}{Ariana Bruno}, \bibinfo{person}{Justin Hu},
  \bibinfo{person}{Brian Ritchken}, \bibinfo{person}{Brendon Jackson},
  \bibinfo{person}{Kelvin Hu}, \bibinfo{person}{Meghna Pancholi},
  \bibinfo{person}{Yuan He}, \bibinfo{person}{Brett Clancy},
  \bibinfo{person}{Chris Colen}, \bibinfo{person}{Fukang Wen},
  \bibinfo{person}{Catherine Leung}, \bibinfo{person}{Siyuan Wang},
  \bibinfo{person}{Leon Zaruvinsky}, \bibinfo{person}{Mateo Espinosa},
  \bibinfo{person}{Rick Lin}, \bibinfo{person}{Zhongling Liu},
  \bibinfo{person}{Jake Padilla}, {and} \bibinfo{person}{Christina
  Delimitrou}.} \bibinfo{year}{2019}\natexlab{a}.
\newblock \showarticletitle{An {Open}-{Source} {Benchmark} {Suite} for
  {Microservices} and {Their} {Hardware}-{Software} {Implications} for {Cloud}
  \& {Edge} {Systems}}. In \bibinfo{booktitle}{\emph{Proceedings of the
  {Twenty}-{Fourth} {International} {Conference} on {Architectural} {Support}
  for {Programming} {Languages} and {Operating} {Systems}}}
  \emph{(\bibinfo{series}{{ASPLOS} '19})}. \bibinfo{publisher}{Association for
  Computing Machinery}, \bibinfo{address}{Providence, RI, USA},
  \bibinfo{pages}{3--18}.
\newblock
\showISBNx{978-1-4503-6240-5}
\urldef\tempurl%
\url{https://doi.org/10.1145/3297858.3304013}
\showDOI{\tempurl}


\bibitem[Gan et~al\mbox{.}(2019b)]%
        {gan-seer-2019}
\bibfield{author}{\bibinfo{person}{Yu Gan}, \bibinfo{person}{Yanqi Zhang},
  \bibinfo{person}{Kelvin Hu}, \bibinfo{person}{Dailun Cheng},
  \bibinfo{person}{Yuan He}, \bibinfo{person}{Meghna Pancholi}, {and}
  \bibinfo{person}{Christina Delimitrou}.} \bibinfo{year}{2019}\natexlab{b}.
\newblock \showarticletitle{Seer: {Leveraging} {Big} {Data} to {Navigate} the
  {Complexity} of {Performance} {Debugging} in {Cloud} {Microservices}}. In
  \bibinfo{booktitle}{\emph{Proceedings of the {Twenty}-{Fourth}
  {International} {Conference} on {Architectural} {Support} for {Programming}
  {Languages} and {Operating} {Systems}}} \emph{(\bibinfo{series}{{ASPLOS}
  '19})}. \bibinfo{publisher}{Association for Computing Machinery},
  \bibinfo{address}{New York, NY, USA}, \bibinfo{pages}{19--33}.
\newblock
\showISBNx{978-1-4503-6240-5}
\urldef\tempurl%
\url{https://doi.org/10.1145/3297858.3304004}
\showDOI{\tempurl}


\bibitem[Garefalakis et~al\mbox{.}(2018)]%
        {garefalakis-medea-2018}
\bibfield{author}{\bibinfo{person}{Panagiotis Garefalakis},
  \bibinfo{person}{Konstantinos Karanasos}, \bibinfo{person}{Peter Pietzuch},
  \bibinfo{person}{Arun Suresh}, {and} \bibinfo{person}{Sriram Rao}.}
  \bibinfo{year}{2018}\natexlab{}.
\newblock \showarticletitle{Medea: scheduling of long running applications in
  shared production clusters}. In \bibinfo{booktitle}{\emph{Proceedings of the
  {Thirteenth} {EuroSys} {Conference}}} \emph{(\bibinfo{series}{{EuroSys}
  '18})}. \bibinfo{publisher}{Association for Computing Machinery},
  \bibinfo{address}{New York, NY, USA}, \bibinfo{pages}{1--13}.
\newblock
\showISBNx{978-1-4503-5584-1}
\urldef\tempurl%
\url{https://doi.org/10.1145/3190508.3190549}
\showDOI{\tempurl}


\bibitem[Gluck(2020)]%
        {gluck-introducing-2020}
\bibfield{author}{\bibinfo{person}{Adam Gluck}.}
  \bibinfo{year}{2020}\natexlab{}.
\newblock \bibinfo{title}{Introducing {Domain}-{Oriented} {Microservice}
  {Architecture}}.
\newblock
\newblock
\urldef\tempurl%
\url{https://eng.uber.com/microservice-architecture/}
\showURL{%
\tempurl}


\bibitem[Guo et~al\mbox{.}(2019)]%
        {guo-who-2019}
\bibfield{author}{\bibinfo{person}{Jing Guo}, \bibinfo{person}{Zihao Chang},
  \bibinfo{person}{Sa Wang}, \bibinfo{person}{Haiyang Ding},
  \bibinfo{person}{Yihui Feng}, \bibinfo{person}{Liang Mao}, {and}
  \bibinfo{person}{Yungang Bao}.} \bibinfo{year}{2019}\natexlab{}.
\newblock \showarticletitle{Who limits the resource efficiency of my
  datacenter: an analysis of {Alibaba} datacenter traces}. In
  \bibinfo{booktitle}{\emph{Proceedings of the {International} {Symposium} on
  {Quality} of {Service}}} \emph{(\bibinfo{series}{{IWQoS} '19})}.
  \bibinfo{publisher}{Association for Computing Machinery},
  \bibinfo{address}{New York, NY, USA}, \bibinfo{pages}{1--10}.
\newblock
\showISBNx{978-1-4503-6778-3}
\urldef\tempurl%
\url{https://doi.org/10.1145/3326285.3329074}
\showDOI{\tempurl}


\bibitem[Gupta et~al\mbox{.}(2020)]%
        {gupta-deeprecsys-2020}
\bibfield{author}{\bibinfo{person}{Udit Gupta}, \bibinfo{person}{Samuel Hsia},
  \bibinfo{person}{Vikram Saraph}, \bibinfo{person}{Xiaodong Wang},
  \bibinfo{person}{Brandon Reagen}, \bibinfo{person}{Gu-Yeon Wei},
  \bibinfo{person}{Hsien-Hsin~S. Lee}, \bibinfo{person}{David Brooks}, {and}
  \bibinfo{person}{Carole-Jean Wu}.} \bibinfo{year}{2020}\natexlab{}.
\newblock \showarticletitle{{DeepRecSys}: {A} {System} for {Optimizing}
  {End}-{To}-{End} {At}-{Scale} {Neural} {Recommendation} {Inference}}. In
  \bibinfo{booktitle}{\emph{2020 {ACM}/{IEEE} 47th {Annual} {International}
  {Symposium} on {Computer} {Architecture} ({ISCA})}}.
  \bibinfo{pages}{982--995}.
\newblock
\urldef\tempurl%
\url{https://doi.org/10.1109/ISCA45697.2020.00084}
\showDOI{\tempurl}


\bibitem[Haarnoja et~al\mbox{.}(2018)]%
        {haarnoja-soft-2018}
\bibfield{author}{\bibinfo{person}{Tuomas Haarnoja}, \bibinfo{person}{Aurick
  Zhou}, \bibinfo{person}{Pieter Abbeel}, {and} \bibinfo{person}{Sergey
  Levine}.} \bibinfo{year}{2018}\natexlab{}.
\newblock \showarticletitle{Soft {Actor}-{Critic}: {Off}-{Policy} {Maximum}
  {Entropy} {Deep} {Reinforcement} {Learning} with a {Stochastic} {Actor}}.
\newblock \bibinfo{journal}{\emph{arXiv:1801.01290 [cs, stat]}}
  (\bibinfo{date}{Aug.} \bibinfo{year}{2018}).
\newblock
\urldef\tempurl%
\url{http://arxiv.org/abs/1801.01290}
\showURL{%
\tempurl}
\newblock
\shownote{arXiv: 1801.01290}.


\bibitem[Haarnoja et~al\mbox{.}(2019)]%
        {haarnoja-soft-2019}
\bibfield{author}{\bibinfo{person}{Tuomas Haarnoja}, \bibinfo{person}{Aurick
  Zhou}, \bibinfo{person}{Kristian Hartikainen}, \bibinfo{person}{George
  Tucker}, \bibinfo{person}{Sehoon Ha}, \bibinfo{person}{Jie Tan},
  \bibinfo{person}{Vikash Kumar}, \bibinfo{person}{Henry Zhu},
  \bibinfo{person}{Abhishek Gupta}, \bibinfo{person}{Pieter Abbeel}, {and}
  \bibinfo{person}{Sergey Levine}.} \bibinfo{year}{2019}\natexlab{}.
\newblock \showarticletitle{Soft {Actor}-{Critic} {Algorithms} and
  {Applications}}.
\newblock \bibinfo{journal}{\emph{arXiv:1812.05905 [cs, stat]}}
  (\bibinfo{date}{Jan.} \bibinfo{year}{2019}).
\newblock
\urldef\tempurl%
\url{http://arxiv.org/abs/1812.05905}
\showURL{%
\tempurl}
\newblock
\shownote{arXiv: 1812.05905}.


\bibitem[Hindman et~al\mbox{.}(2011)]%
        {hindman-mesos-2011}
\bibfield{author}{\bibinfo{person}{Benjamin Hindman}, \bibinfo{person}{Andy
  Konwinski}, \bibinfo{person}{Matei Zaharia}, \bibinfo{person}{Ali Ghodsi},
  \bibinfo{person}{Anthony~D. Joseph}, \bibinfo{person}{Randy Katz},
  \bibinfo{person}{Scott Shenker}, {and} \bibinfo{person}{Ion Stoica}.}
  \bibinfo{year}{2011}\natexlab{}.
\newblock \showarticletitle{Mesos: a platform for fine-grained resource sharing
  in the data center}. In \bibinfo{booktitle}{\emph{Proceedings of the 8th
  {USENIX} conference on {Networked} systems design and implementation}}
  \emph{(\bibinfo{series}{{NSDI}'11})}. \bibinfo{publisher}{USENIX
  Association}, \bibinfo{address}{USA}, \bibinfo{pages}{295--308}.
\newblock


\bibitem[Hou et~al\mbox{.}(2021)]%
        {hou-alphar-2021}
\bibfield{author}{\bibinfo{person}{Xiaofeng Hou}, \bibinfo{person}{Chao Li},
  \bibinfo{person}{Jiacheng Liu}, \bibinfo{person}{Lu Zhang},
  \bibinfo{person}{Shaolei Ren}, \bibinfo{person}{Jingwen Leng},
  \bibinfo{person}{Quan Chen}, {and} \bibinfo{person}{Minyi Guo}.}
  \bibinfo{year}{2021}\natexlab{}.
\newblock \showarticletitle{{AlphaR}: {Learning}-{Powered} {Resource}
  {Management} for {Irregular}, {Dynamic} {Microservice} {Graph}}. In
  \bibinfo{booktitle}{\emph{2021 {IEEE} {International} {Parallel} and
  {Distributed} {Processing} {Symposium} ({IPDPS})}}.
  \bibinfo{pages}{797--806}.
\newblock
\urldef\tempurl%
\url{https://doi.org/10.1109/IPDPS49936.2021.00089}
\showDOI{\tempurl}
\newblock
\shownote{ISSN: 1530-2075}.


\bibitem[Jamshidi et~al\mbox{.}(2018)]%
        {jamshidi-microservices-2018}
\bibfield{author}{\bibinfo{person}{Pooyan Jamshidi}, \bibinfo{person}{Claus
  Pahl}, \bibinfo{person}{Nabor~C. Mendonça}, \bibinfo{person}{James Lewis},
  {and} \bibinfo{person}{Stefan Tilkov}.} \bibinfo{year}{2018}\natexlab{}.
\newblock \showarticletitle{Microservices: {The} {Journey} {So} {Far} and
  {Challenges} {Ahead}}.
\newblock \bibinfo{journal}{\emph{IEEE Software}} \bibinfo{volume}{35},
  \bibinfo{number}{3} (\bibinfo{date}{May} \bibinfo{year}{2018}),
  \bibinfo{pages}{24--35}.
\newblock
\showISSN{1937-4194}
\urldef\tempurl%
\url{https://doi.org/10.1109/MS.2018.2141039}
\showDOI{\tempurl}
\newblock
\shownote{Conference Name: IEEE Software}.


\bibitem[Kulkarni et~al\mbox{.}(2018)]%
        {kulkarni-leveraging-2018}
\bibfield{author}{\bibinfo{person}{Neeraj Kulkarni}, \bibinfo{person}{Feng Qi},
  {and} \bibinfo{person}{Christina Delimitrou}.}
  \bibinfo{year}{2018}\natexlab{}.
\newblock \showarticletitle{Leveraging {Approximation} to {Improve}
  {Datacenter} {Resource} {Efficiency}}.
\newblock \bibinfo{journal}{\emph{IEEE Computer Architecture Letters}}
  \bibinfo{volume}{17}, \bibinfo{number}{2} (\bibinfo{date}{July}
  \bibinfo{year}{2018}), \bibinfo{pages}{171--174}.
\newblock
\showISSN{1556-6064}
\urldef\tempurl%
\url{https://doi.org/10.1109/LCA.2018.2845841}
\showDOI{\tempurl}
\newblock
\shownote{Conference Name: IEEE Computer Architecture Letters}.


\bibitem[Kwan et~al\mbox{.}(2019)]%
        {HyScale}
\bibfield{author}{\bibinfo{person}{Anthony Kwan}, \bibinfo{person}{Jonathon
  Wong}, \bibinfo{person}{Hans-Arno Jacobsen}, {and} \bibinfo{person}{Vinod
  Muthusamy}.} \bibinfo{year}{2019}\natexlab{}.
\newblock \showarticletitle{HyScale: Hybrid and Network Scaling of Dockerized
  Microservices in Cloud Data Centres}. In \bibinfo{booktitle}{\emph{2019 IEEE
  39th International Conference on Distributed Computing Systems (ICDCS)}}.
  \bibinfo{pages}{80--90}.
\newblock
\urldef\tempurl%
\url{https://doi.org/10.1109/ICDCS.2019.00017}
\showDOI{\tempurl}


\bibitem[Li et~al\mbox{.}(2018)]%
        {li-model-free-2018}
\bibfield{author}{\bibinfo{person}{Teng Li}, \bibinfo{person}{Zhiyuan Xu},
  \bibinfo{person}{Jian Tang}, {and} \bibinfo{person}{Yanzhi Wang}.}
  \bibinfo{year}{2018}\natexlab{}.
\newblock \showarticletitle{Model-free control for distributed stream data
  processing using deep reinforcement learning}.
\newblock \bibinfo{journal}{\emph{Proceedings of the VLDB Endowment}}
  \bibinfo{volume}{11}, \bibinfo{number}{6} (\bibinfo{date}{Feb.}
  \bibinfo{year}{2018}), \bibinfo{pages}{705--718}.
\newblock
\showISSN{2150-8097}
\urldef\tempurl%
\url{https://doi.org/10.14778/3199517.3199521}
\showDOI{\tempurl}


\bibitem[Li et~al\mbox{.}(2020)]%
        {li-dynamic-2020}
\bibfield{author}{\bibinfo{person}{Yuhao Li}, \bibinfo{person}{Dan Sun}, {and}
  \bibinfo{person}{Benjamin~C. Lee}.} \bibinfo{year}{2020}\natexlab{}.
\newblock \showarticletitle{Dynamic {Colocation} {Policies} with
  {Reinforcement} {Learning}}.
\newblock \bibinfo{journal}{\emph{ACM Transactions on Architecture and Code
  Optimization}} \bibinfo{volume}{17}, \bibinfo{number}{1}
  (\bibinfo{date}{March} \bibinfo{year}{2020}), \bibinfo{pages}{1:1--1:25}.
\newblock
\showISSN{1544-3566}
\urldef\tempurl%
\url{https://doi.org/10.1145/3375714}
\showDOI{\tempurl}


\bibitem[Lin et~al\mbox{.}(2019)]%
        {lin-ant-2019}
\bibfield{author}{\bibinfo{person}{Miao Lin}, \bibinfo{person}{Jianqing Xi},
  \bibinfo{person}{Weihua Bai}, {and} \bibinfo{person}{Jiayin Wu}.}
  \bibinfo{year}{2019}\natexlab{}.
\newblock \showarticletitle{Ant {Colony} {Algorithm} for {Multi}-{Objective}
  {Optimization} of {Container}-{Based} {Microservice} {Scheduling} in
  {Cloud}}.
\newblock \bibinfo{journal}{\emph{IEEE Access}}  \bibinfo{volume}{7}
  (\bibinfo{year}{2019}), \bibinfo{pages}{83088--83100}.
\newblock
\showISSN{2169-3536}
\urldef\tempurl%
\url{https://doi.org/10.1109/ACCESS.2019.2924414}
\showDOI{\tempurl}
\newblock
\shownote{Conference Name: IEEE Access}.


\bibitem[Linden(2006)]%
        {linden-geeking-2006}
\bibfield{author}{\bibinfo{person}{Greg Linden}.}
  \bibinfo{year}{2006}\natexlab{}.
\newblock \bibinfo{title}{Geeking with {Greg}: {Slides} from my talk at
  {Stanford}}.
\newblock
\newblock
\urldef\tempurl%
\url{https://glinden.blogspot.com/2006/12/slides-from-my-talk-at-stanford.html}
\showURL{%
\tempurl}


\bibitem[Lo et~al\mbox{.}(2015)]%
        {lo-heracles-2015}
\bibfield{author}{\bibinfo{person}{David Lo}, \bibinfo{person}{Liqun Cheng},
  \bibinfo{person}{Rama Govindaraju}, \bibinfo{person}{Parthasarathy
  Ranganathan}, {and} \bibinfo{person}{Christos Kozyrakis}.}
  \bibinfo{year}{2015}\natexlab{}.
\newblock \showarticletitle{Heracles: {Improving} resource efficiency at
  scale}. In \bibinfo{booktitle}{\emph{2015 {ACM}/{IEEE} 42nd {Annual}
  {International} {Symposium} on {Computer} {Architecture} ({ISCA})}}.
  \bibinfo{pages}{450--462}.
\newblock
\urldef\tempurl%
\url{https://doi.org/10.1145/2749469.2749475}
\showDOI{\tempurl}
\newblock
\shownote{ISSN: 1063-6897}.


\bibitem[Lundberg and Lee(2017)]%
        {lundberg-unified-2017}
\bibfield{author}{\bibinfo{person}{Scott~M Lundberg} {and}
  \bibinfo{person}{Su-In Lee}.} \bibinfo{year}{2017}\natexlab{}.
\newblock \showarticletitle{A {Unified} {Approach} to {Interpreting} {Model}
  {Predictions}}.
\newblock In \bibinfo{booktitle}{\emph{Advances in {Neural} {Information}
  {Processing} {Systems} 30}}, \bibfield{editor}{\bibinfo{person}{I.~Guyon},
  \bibinfo{person}{U.~V. Luxburg}, \bibinfo{person}{S.~Bengio},
  \bibinfo{person}{H.~Wallach}, \bibinfo{person}{R.~Fergus},
  \bibinfo{person}{S.~Vishwanathan}, {and} \bibinfo{person}{R.~Garnett}}
  (Eds.). \bibinfo{publisher}{Curran Associates, Inc.},
  \bibinfo{pages}{4765--4774}.
\newblock
\urldef\tempurl%
\url{http://papers.nips.cc/paper/7062-a-unified-approach-to-interpreting-model-predictions.pdf}
\showURL{%
\tempurl}


\bibitem[Mao et~al\mbox{.}(2017)]%
        {mao-neural-2017}
\bibfield{author}{\bibinfo{person}{Hongzi Mao}, \bibinfo{person}{Ravi
  Netravali}, {and} \bibinfo{person}{Mohammad Alizadeh}.}
  \bibinfo{year}{2017}\natexlab{}.
\newblock \showarticletitle{Neural {Adaptive} {Video} {Streaming} with
  {Pensieve}}. In \bibinfo{booktitle}{\emph{Proceedings of the {Conference} of
  the {ACM} {Special} {Interest} {Group} on {Data} {Communication}}}
  \emph{(\bibinfo{series}{{SIGCOMM} '17})}. \bibinfo{publisher}{Association for
  Computing Machinery}, \bibinfo{address}{New York, NY, USA},
  \bibinfo{pages}{197--210}.
\newblock
\showISBNx{978-1-4503-4653-5}
\urldef\tempurl%
\url{https://doi.org/10.1145/3098822.3098843}
\showDOI{\tempurl}


\bibitem[Mauro(2015)]%
        {mauro-microservices-2015}
\bibfield{author}{\bibinfo{person}{Tony Mauro}.}
  \bibinfo{year}{2015}\natexlab{}.
\newblock \bibinfo{title}{Microservices at {Netflix}: {Lessons} for
  {Architectural} {Design}}.
\newblock
\newblock
\urldef\tempurl%
\url{https://www.nginx.com/blog/microservices-at-netflix-architectural-best-practices/}
\showURL{%
\tempurl}


\bibitem[Mnih et~al\mbox{.}(2016)]%
        {mnih-asynchronous-2016}
\bibfield{author}{\bibinfo{person}{Volodymyr Mnih},
  \bibinfo{person}{Adria~Puigdomenech Badia}, \bibinfo{person}{Mehdi Mirza},
  \bibinfo{person}{Alex Graves}, \bibinfo{person}{Timothy Lillicrap},
  \bibinfo{person}{Tim Harley}, \bibinfo{person}{David Silver}, {and}
  \bibinfo{person}{Koray Kavukcuoglu}.} \bibinfo{year}{2016}\natexlab{}.
\newblock \showarticletitle{Asynchronous {Methods} for {Deep} {Reinforcement}
  {Learning}}. In \bibinfo{booktitle}{\emph{International {Conference} on
  {Machine} {Learning}}}. \bibinfo{pages}{1928--1937}.
\newblock
\urldef\tempurl%
\url{http://proceedings.mlr.press/v48/mniha16.html}
\showURL{%
\tempurl}


\bibitem[Mnih et~al\mbox{.}(2015)]%
        {mnih-human-level-2015}
\bibfield{author}{\bibinfo{person}{Volodymyr Mnih}, \bibinfo{person}{Koray
  Kavukcuoglu}, \bibinfo{person}{David Silver}, \bibinfo{person}{Andrei~A.
  Rusu}, \bibinfo{person}{Joel Veness}, \bibinfo{person}{Marc~G. Bellemare},
  \bibinfo{person}{Alex Graves}, \bibinfo{person}{Martin Riedmiller},
  \bibinfo{person}{Andreas~K. Fidjeland}, \bibinfo{person}{Georg Ostrovski},
  \bibinfo{person}{Stig Petersen}, \bibinfo{person}{Charles Beattie},
  \bibinfo{person}{Amir Sadik}, \bibinfo{person}{Ioannis Antonoglou},
  \bibinfo{person}{Helen King}, \bibinfo{person}{Dharshan Kumaran},
  \bibinfo{person}{Daan Wierstra}, \bibinfo{person}{Shane Legg}, {and}
  \bibinfo{person}{Demis Hassabis}.} \bibinfo{year}{2015}\natexlab{}.
\newblock \showarticletitle{Human-level control through deep reinforcement
  learning}.
\newblock \bibinfo{journal}{\emph{Nature}} \bibinfo{volume}{518},
  \bibinfo{number}{7540} (\bibinfo{date}{Feb.} \bibinfo{year}{2015}),
  \bibinfo{pages}{529--533}.
\newblock
\showISSN{1476-4687}
\urldef\tempurl%
\url{https://doi.org/10.1038/nature14236}
\showDOI{\tempurl}


\bibitem[Nishtala et~al\mbox{.}(2017)]%
        {nishtala-hipster-2017}
\bibfield{author}{\bibinfo{person}{Rajiv Nishtala}, \bibinfo{person}{Paul
  Carpenter}, \bibinfo{person}{Vinicius Petrucci}, {and}
  \bibinfo{person}{Xavier Martorell}.} \bibinfo{year}{2017}\natexlab{}.
\newblock \showarticletitle{The {Hipster} {Approach} for {Improving} {Cloud}
  {System} {Efficiency}}.
\newblock \bibinfo{journal}{\emph{ACM Transactions on Computer Systems}}
  \bibinfo{volume}{35}, \bibinfo{number}{3} (\bibinfo{date}{Dec.}
  \bibinfo{year}{2017}), \bibinfo{pages}{8:1--8:28}.
\newblock
\showISSN{0734-2071}
\urldef\tempurl%
\url{https://doi.org/10.1145/3144168}
\showDOI{\tempurl}


\bibitem[Park et~al\mbox{.}(2021)]%
        {park-graf-2021}
\bibfield{author}{\bibinfo{person}{Jinwoo Park}, \bibinfo{person}{Byungkwon
  Choi}, \bibinfo{person}{Chunghan Lee}, {and} \bibinfo{person}{Dongsu Han}.}
  \bibinfo{year}{2021}\natexlab{}.
\newblock \showarticletitle{{GRAF}: a graph neural network based proactive
  resource allocation framework for {SLO}-oriented microservices}. In
  \bibinfo{booktitle}{\emph{Proceedings of the 17th {International}
  {Conference} on emerging {Networking} {EXperiments} and {Technologies}}}
  \emph{(\bibinfo{series}{{CoNEXT} '21})}. \bibinfo{publisher}{Association for
  Computing Machinery}, \bibinfo{address}{New York, NY, USA},
  \bibinfo{pages}{154--167}.
\newblock
\showISBNx{978-1-4503-9098-9}
\urldef\tempurl%
\url{https://doi.org/10.1145/3485983.3494866}
\showDOI{\tempurl}


\bibitem[Patel and Tiwari(2020)]%
        {patel-clite-2020}
\bibfield{author}{\bibinfo{person}{Tirthak Patel} {and} \bibinfo{person}{Devesh
  Tiwari}.} \bibinfo{year}{2020}\natexlab{}.
\newblock \showarticletitle{{CLITE}: {Efficient} and {QoS}-{Aware}
  {Co}-{Location} of {Multiple} {Latency}-{Critical} {Jobs} for {Warehouse}
  {Scale} {Computers}}. In \bibinfo{booktitle}{\emph{2020 {IEEE}
  {International} {Symposium} on {High} {Performance} {Computer} {Architecture}
  ({HPCA})}}. \bibinfo{pages}{193--206}.
\newblock
\urldef\tempurl%
\url{https://doi.org/10.1109/HPCA47549.2020.00025}
\showDOI{\tempurl}
\newblock
\shownote{ISSN: 2378-203X}.


\bibitem[Qiu et~al\mbox{.}(2020)]%
        {qiu-firm-2020}
\bibfield{author}{\bibinfo{person}{Haoran Qiu}, \bibinfo{person}{Subho~S.
  Banerjee}, \bibinfo{person}{Saurabh Jha}, \bibinfo{person}{Zbigniew~T.
  Kalbarczyk}, {and} \bibinfo{person}{Ravishankar~K. Iyer}.}
  \bibinfo{year}{2020}\natexlab{}.
\newblock \showarticletitle{Firm: {An} intelligent fine-grained resource
  management framework for {SLO}-{Oriented} microservices}. In
  \bibinfo{booktitle}{\emph{Proceedings of the 14th {USENIX} {Symposium} on
  {Operating} {Systems} {Design} and {Implementation}, {OSDI} 2020}}.
  \bibinfo{publisher}{USENIX Association}, \bibinfo{pages}{805--825}.
\newblock
\urldef\tempurl%
\url{https://experts.illinois.edu/en/publications/firm-an-intelligent-fine-grained-resource-management-framework-fo-2}
\showURL{%
\tempurl}


\bibitem[Rzadca et~al\mbox{.}(2020)]%
        {rzadca-autopilot-2020}
\bibfield{author}{\bibinfo{person}{Krzysztof Rzadca}, \bibinfo{person}{Paweł
  Findeisen}, \bibinfo{person}{Jacek Świderski}, \bibinfo{person}{Przemyslaw
  Zych}, \bibinfo{person}{Przemyslaw Broniek}, \bibinfo{person}{Jarek
  Kusmierek}, \bibinfo{person}{Paweł~Krzysztof Nowak}, \bibinfo{person}{Beata
  Strack}, \bibinfo{person}{Piotr Witusowski}, \bibinfo{person}{Steven Hand},
  {and} \bibinfo{person}{John Wilkes}.} \bibinfo{year}{2020}\natexlab{}.
\newblock \showarticletitle{Autopilot: {Workload} {Autoscaling} at {Google}
  {Scale}}. In \bibinfo{booktitle}{\emph{Proceedings of the {Fifteenth}
  {European} {Conference} on {Computer} {Systems}}}.
\newblock
\urldef\tempurl%
\url{https://dl.acm.org/doi/10.1145/3342195.3387524}
\showURL{%
\tempurl}


\bibitem[Satnic(2021)]%
        {satnic-amazon-nodate}
\bibfield{author}{\bibinfo{person}{Cristian Satnic}.}
  \bibinfo{year}{2021}\natexlab{}.
\newblock \bibinfo{title}{Amazon, microservices and the birth of {AWS} cloud
  computing {\textbar} {LinkedIn}}.
\newblock
\newblock
\urldef\tempurl%
\url{https://www.linkedin.com/pulse/amazon-microservices-birth-aws-cloud-computing-cristian-satnic/}
\showURL{%
\tempurl}


\bibitem[Thamsen et~al\mbox{.}(2017)]%
        {thamsen-learning-2017}
\bibfield{author}{\bibinfo{person}{L. Thamsen}, \bibinfo{person}{I.
  Verbitskiy}, \bibinfo{person}{Benjamin Rabier}, {and} \bibinfo{person}{O.
  Kao}.} \bibinfo{year}{2017}\natexlab{}.
\newblock \showarticletitle{Learning {Efficient} {Co}-locations for
  {Scheduling} {Distributed} {Dataflows} in {Shared} {Clusters}}.
\newblock


\bibitem[Xu et~al\mbox{.}(2022)]%
        {CoScal}
\bibfield{author}{\bibinfo{person}{Minxian Xu}, \bibinfo{person}{Chenghao
  Song}, \bibinfo{person}{Shashikant Ilager}, \bibinfo{person}{Sukhpal~Singh
  Gill}, \bibinfo{person}{Juanjuan Zhao}, \bibinfo{person}{Kejiang Ye}, {and}
  \bibinfo{person}{Chengzhong Xu}.} \bibinfo{year}{2022}\natexlab{}.
\newblock \showarticletitle{CoScal: Multi-faceted Scaling of Microservices with
  Reinforcement Learning}.
\newblock \bibinfo{journal}{\emph{IEEE Transactions on Network and Service
  Management}} (\bibinfo{year}{2022}), \bibinfo{pages}{1--1}.
\newblock
\urldef\tempurl%
\url{https://doi.org/10.1109/TNSM.2022.3210211}
\showDOI{\tempurl}


\bibitem[Xu et~al\mbox{.}(2018)]%
        {xu-experience-driven-2018}
\bibfield{author}{\bibinfo{person}{Zhiyuan Xu}, \bibinfo{person}{Jian Tang},
  \bibinfo{person}{Jingsong Meng}, \bibinfo{person}{Weiyi Zhang},
  \bibinfo{person}{Yanzhi Wang}, \bibinfo{person}{Chi~Harold Liu}, {and}
  \bibinfo{person}{Dejun Yang}.} \bibinfo{year}{2018}\natexlab{}.
\newblock \showarticletitle{Experience-driven {Networking}: {A} {Deep}
  {Reinforcement} {Learning} based {Approach}}. In
  \bibinfo{booktitle}{\emph{{IEEE} {INFOCOM} 2018 - {IEEE} {Conference} on
  {Computer} {Communications}}}. \bibinfo{pages}{1871--1879}.
\newblock
\urldef\tempurl%
\url{https://doi.org/10.1109/INFOCOM.2018.8485853}
\showDOI{\tempurl}


\bibitem[Yang et~al\mbox{.}(2013)]%
        {yang-bubble-flux-2013}
\bibfield{author}{\bibinfo{person}{Hailong Yang}, \bibinfo{person}{Alex
  Breslow}, \bibinfo{person}{Jason Mars}, {and} \bibinfo{person}{Lingjia
  Tang}.} \bibinfo{year}{2013}\natexlab{}.
\newblock \showarticletitle{Bubble-flux: precise online {QoS} management for
  increased utilization in warehouse scale computers}.
\newblock \bibinfo{journal}{\emph{ACM SIGARCH Computer Architecture News}}
  \bibinfo{volume}{41}, \bibinfo{number}{3} (\bibinfo{date}{June}
  \bibinfo{year}{2013}), \bibinfo{pages}{607--618}.
\newblock
\showISSN{0163-5964}
\urldef\tempurl%
\url{https://doi.org/10.1145/2508148.2485974}
\showDOI{\tempurl}


\bibitem[Yang et~al\mbox{.}(2017)]%
        {yang-powerchief-2017}
\bibfield{author}{\bibinfo{person}{Hailong Yang}, \bibinfo{person}{Quan Chen},
  \bibinfo{person}{Moeiz Riaz}, \bibinfo{person}{Zhongzhi Luan},
  \bibinfo{person}{Lingjia Tang}, {and} \bibinfo{person}{Jason Mars}.}
  \bibinfo{year}{2017}\natexlab{}.
\newblock \showarticletitle{{PowerChief}: {Intelligent} {Power} {Allocation}
  for {Multi}-{Stage} {Applications} to {Improve} {Responsiveness} on {Power}
  {Constrained} {CMP}}. In \bibinfo{booktitle}{\emph{Proceedings of the 44th
  {Annual} {International} {Symposium} on {Computer} {Architecture}}}
  \emph{(\bibinfo{series}{{ISCA} '17})}. \bibinfo{publisher}{Association for
  Computing Machinery}, \bibinfo{address}{New York, NY, USA},
  \bibinfo{pages}{133--146}.
\newblock
\showISBNx{978-1-4503-4892-8}
\urldef\tempurl%
\url{https://doi.org/10.1145/3079856.3080224}
\showDOI{\tempurl}


\bibitem[Yang et~al\mbox{.}(2019)]%
        {yang-miras-2019}
\bibfield{author}{\bibinfo{person}{Zhe Yang}, \bibinfo{person}{Phuong Nguyen},
  \bibinfo{person}{Haiming Jin}, {and} \bibinfo{person}{Klara Nahrstedt}.}
  \bibinfo{year}{2019}\natexlab{}.
\newblock \showarticletitle{{MIRAS}: {Model}-based {Reinforcement} {Learning}
  for {Microservice} {Resource} {Allocation} over {Scientific} {Workflows}}. In
  \bibinfo{booktitle}{\emph{2019 {IEEE} 39th {International} {Conference} on
  {Distributed} {Computing} {Systems} ({ICDCS})}}. \bibinfo{pages}{122--132}.
\newblock
\urldef\tempurl%
\url{https://doi.org/10.1109/ICDCS.2019.00021}
\showDOI{\tempurl}
\newblock
\shownote{ISSN: 2575-8411}.


\bibitem[Yu et~al\mbox{.}(2022)]%
        {IoTDRL}
\bibfield{author}{\bibinfo{person}{Yinbo Yu}, \bibinfo{person}{Jiajia Liu},
  {and} \bibinfo{person}{Jing Fang}.} \bibinfo{year}{2022}\natexlab{}.
\newblock \showarticletitle{Online Microservice Orchestration for IoT via
  Multiobjective Deep Reinforcement Learning}.
\newblock \bibinfo{journal}{\emph{IEEE Internet of Things Journal}}
  \bibinfo{volume}{9}, \bibinfo{number}{18} (\bibinfo{year}{2022}),
  \bibinfo{pages}{17513--17525}.
\newblock
\urldef\tempurl%
\url{https://doi.org/10.1109/JIOT.2022.3155598}
\showDOI{\tempurl}


\bibitem[Zhang et~al\mbox{.}(2021)]%
        {zhang-sinan-2021}
\bibfield{author}{\bibinfo{person}{Yanqi Zhang}, \bibinfo{person}{Weizhe Hua},
  \bibinfo{person}{Zhuangzhuang Zhou}, \bibinfo{person}{G.~Edward Suh}, {and}
  \bibinfo{person}{Christina Delimitrou}.} \bibinfo{year}{2021}\natexlab{}.
\newblock \showarticletitle{Sinan: {ML}-based and {QoS}-aware resource
  management for cloud microservices}. In \bibinfo{booktitle}{\emph{Proceedings
  of the 26th {ACM} {International} {Conference} on {Architectural} {Support}
  for {Programming} {Languages} and {Operating} {Systems}}}
  \emph{(\bibinfo{series}{{ASPLOS} 2021})}. \bibinfo{publisher}{Association for
  Computing Machinery}, \bibinfo{address}{New York, NY, USA},
  \bibinfo{pages}{167--181}.
\newblock
\showISBNx{978-1-4503-8317-2}
\urldef\tempurl%
\url{https://doi.org/10.1145/3445814.3446693}
\showDOI{\tempurl}


\end{thebibliography}


\end{document}